\documentclass[draftcls, 12pt, onecolumn, twoside]{IEEEtran}

\usepackage{amsfonts}
\usepackage{amsmath}
\usepackage{amssymb}
\usepackage{lscape}
\usepackage{epsf}
\usepackage{graphicx}
\usepackage{psfrag}
\usepackage{color}

\usepackage[nolist,printonlyused]{acronym}

\acrodef{TOA}{time-of-arrival} \acrodef{UWB}{ultrawide-band}
\acrodef{WB}{wideband} \acrodef{LOS}{line-of-sight}
\acrodef{NB}{narrow band} \acrodef{RCS}{radar cross section}
\acrodef{PRF}{pulse repetition frequency} \acrodef{r.v.}{random
variable} \acrodef{i.i.d.}{independent, identically distributed}
\acrodef{p.d.f.}{probability distribution function}
\acrodef{c.d.f.}{cumulative distribution function}
\acrodef{ch.f.}{characteristic function} \acrodef{AWGN}{additive
white gaussian noise} \acrodef{BSC}{binary symmetric channels}
\acrodef{BPSK}{binary phase shift keying}
\acrodef{SNR}{signal-to-noise ratio}
\acrodef{SIR}{signal-to-interference ratio}
\acrodef{CCI}{co-channel interference} \acrodef{CSI}{channel state
information} \acrodef{MIMO}{multiple-input/multiple-output}
\acrodef{RADAR}{radio detection and ranging}
\acrodef{CRB}{Cramer-Rao bound} \acrodef{CN}{check node}
\acrodef{VN}{variable node} \acrodef{CND}{check node decoder}
\acrodef{VND}{variable node decoder} \acrodef{GLDPC}{generalized
LDPC} \acrodef{LDPC}{low-density parity-check}
\acrodef{SPC}{single parity-check}
\acrodef{D-GLDPC}{doubly-generalized LDPC} \acrodef{BEC}{binary
erasure channel} \acrodef{MAP}{maximum \emph{a posteriori}}

\newtheorem{theorem}{\indent Theorem}[section]
\newtheorem{lemma}[theorem]{\indent Lemma}
\newtheorem{corollary}[theorem]{\indent Corollary}

\newtheorem{EXAMPLE}{\indent Example}[section]
\newtheorem{definition}{\indent Definition}[section]

\newcommand{\cA}{{\mathcal{A}}}

\newcommand{\cB}{{\mathcal{B}}}

\newcommand{\cM}{{\mathcal{M}}}

\newcommand{\coeff}{{\mbox{Coeff }}}

\newcommand{\bldbeta}{{\mbox{\boldmath $\beta$}}}
\newcommand{\bldsmallbeta}{{\mbox{\scriptsize \boldmath $\beta$}}}

\newcommand{\bldeta}{{\mbox{\boldmath $\eta$}}}
\newcommand{\bldsmalleta}{{\mbox{\scriptsize \boldmath $\eta$}}}
\newcommand{\bldepsilon}{{\mbox{\boldmath $\epsilon$}}}
\newcommand{\bldsmallepsilon}{{\mbox{\scriptsize \boldmath $\epsilon$}}}

\newcommand{\bldgamma}{{\mbox{\boldmath $\gamma$}}}
\newcommand{\bldsmallgamma}{{\mbox{\scriptsize \boldmath $\gamma$}}}

\newcommand{\bldnu}{{\mbox{\boldmath $\nu$}}}
\newcommand{\bldsmallnu}{{\mbox{\scriptsize \boldmath $\nu$}}}

\newcommand{\bldtheta}{{\mbox{\boldmath $\theta$}}}
\newcommand{\bldsmalltheta}{{\mbox{\scriptsize \boldmath $\theta$}}}

\newcommand{\bldupsilon}{{\mbox{\boldmath $\upsilon$}}}
\newcommand{\bldsmallupsilon}{{\mbox{\scriptsize \boldmath $\upsilon$}}}

\newcommand{\asympequalm}{\stackrel{m}{\approx}}
\newcommand{\asympequaln}{\stackrel{n}{\approx}}

    \def\squarebox#1{\hbox to #1{\hfill\vbox to #1{\vfill}}}

\newlength{\Algwidth}

\title{On the Growth Rate of the Weight Distribution of Irregular Doubly-Generalized LDPC Codes
\thanks{%
    This work was supported in part by the EC under Seventh FP grant agreement ICT OPTIMIX n. INFSO-ICT-214625 and in part by the University of Bologna (ISA-ESRF fellowship). The material in this paper was presented in part at the 46-th International Allerton Conference on Communication, Control and Computing, Monticello, Illinois, September 2008.  
    \newline 
		M. F. Flanagan is with the Department of Electronic and Electrical Engineering, University College Dublin, Belfield, Dublin 4, Ireland (e-mail:mark.flanagan@ieee.org). 
		\newline
    E. Paolini and M. Chiani are with DEIS, University of Bologna, Via Venezia 52, 47023 Cesena (FC), Cesena, Italy (e-mail:e.paolini@unibo.it, marco.chiani@unibo.it).
    \newline
    M. P. C. Fossorier is with ETIS ENSEA, UCP, CNRS UMR-8051, 6 avenue du Ponceau, 95014 Cergy Pontoise, France (e-mail: mfossorier@ieee.org). 
    }        
}

\author{Mark F. Flanagan,~\IEEEmembership{Member,~IEEE,} Enrico Paolini,~\IEEEmembership{Member,~IEEE,} Marco Chiani,~\IEEEmembership{Senior Member,~IEEE,} and~Marc~P.~C.~Fossorier,~\IEEEmembership{Fellow,~IEEE}}
\begin{document}
\maketitle

\begin{abstract}
In this paper, an expression for the asymptotic growth rate of the
number of small linear-weight codewords of irregular doubly-generalized
LDPC (D-GLDPC) codes is derived. The expression is compact and generalizes 
existing results for LDPC and generalized LDPC (GLDPC) codes. Ensembles with check or variable node minimum distance greater than $2$ are shown to be have good growth rate behavior, while for other ensembles a fundamental parameter is identified which discriminates between an asymptotically small and an
asymptotically large expected number of small linear-weight
codewords. Also, in the latter case it is shown that the growth rate depends only on the check and variable nodes with minimum distance $2$.  An important connection between this new result and the stability condition of D-GLDPC codes
over the BEC is highlighted. Such a connection, previously
observed for LDPC and GLDPC codes, is now extended to
the case of D-GLDPC codes. Finally, it is shown that the analysis may be extended to include the growth rate of the stopping set size distribution of irregular D-GLDPC codes. 
\end{abstract}

\section{Introduction}

Recently, \ac{LDPC} codes have been intensively studied due to
their near-Shannon-limit performance under iterative
belief-propagation decoding. Binary regular \ac{LDPC} codes were first
proposed by Gallager in 1963 \cite{Gallager}. In the last
decade the capability of irregular \ac{LDPC} codes to outperform
regular ones in the waterfall region of the performance curve and
to asymptotically approach (or even achieve) the communication
channel capacity has been recognized and deeply investigated (see
for instance
\cite{luby01:improved,luby01:efficient,richardson01:design,richardson01:dB,pfister05:capacity-achieving,pfister07:ara}).

It is usual to represent an \ac{LDPC} code as a bipartite graph,
i.e., as a graph where the nodes are grouped into two disjoint
sets, namely, the \acp{VN} and the \acp{CN}, such that each
edge may only connect a \ac{VN} to a \ac{CN}. The bipartite graph
is also known as a Tanner graph \cite{Tanner_GLDPC}. In the Tanner
graph of an \ac{LDPC} code, a generic degree-$q$ \ac{VN} can
be interpreted as a length-$q$ repetition code, as it repeats $q$
times its single information bit towards the \acp{CN}. Similarly, a degree-$s$ 
\ac{CN} of an \ac{LDPC} code can be interpreted
as a length-$s$ \ac{SPC} code, as it checks the parity of the
$s$ \acp{VN} connected to it.

The growth rate of the weight distribution of Gallager's regular
\ac{LDPC} codes was investigated in \cite{Gallager}. The analysis
demonstrated that, provided that the smallest
\ac{VN} degree is at least 3, the ensemble 
has good growth rate behavior, i.e. a code randomly chosen from the ensemble contains an asymptotically small expected 
number of small linear-weight codewords.

More recently, the study of the weight distribution of binary \ac{LDPC} codes
has been extended to irregular ensembles. Pioneering works in this
area are
\cite{litsyn02:ensembles,Burshtein_Miller,Di_Richardson_Urbanke}.
In \cite{Di_Richardson_Urbanke} a complete solution for the growth rate of the weight
distribution of binary irregular \ac{LDPC} codes was
developed. One of the main results of \cite{Di_Richardson_Urbanke} is a connection
between the expected behavior of the weight distribution of a code
randomly chosen from the ensemble and the parameter
$\lambda'(0)\rho'(1)$, $\lambda(x)$ and $\rho(x)$ being the
edge-perspective \ac{VN} and \ac{CN} degree distributions,
respectively. More specifically, it was shown that for a code randomly
chosen from the ensemble, one can expect an exponentially small
number of small linear-weight codewords if $0 \leq
\lambda'(0)\rho'(1)<1$, and an exponentially large number of small
linear-weight codewords if $\lambda'(0)\rho'(1)>1$.

This result establishes a connection between the statistical
properties of the weight distribution of binary irregular \ac{LDPC} codes and
the stability condition of binary irregular \ac{LDPC} codes over the \ac{BEC}
\cite{luby01:efficient,richardson01:design}. If $q^*$ denotes
the \ac{LDPC} asymptotic iterative decoding threshold over the
\ac{BEC}, the stability condition states that we always have
\begin{align}\label{eq:stability_LDPC}
q^* \leq \left[ \lambda'(0)\rho'(1) \right]^{-1}.
\end{align}

Prior to the rediscovery of \ac{LDPC} codes, binary \ac{GLDPC} codes
were introduced by Tanner in 1981 \cite{Tanner_GLDPC}. A
\ac{GLDPC} code generalizes the concept of an \ac{LDPC} code in that
a degree-$s$ \ac{CN} may in principle be any $(s,h)$ linear block
code, $s$ being the code length and $h$ the code dimension. Such a
\ac{CN} accounts for $s-h$ linearly independent parity-check equations. A \ac{CN}
associated with a linear block code which is not a \ac{SPC} code
is said to be a \emph{generalized \ac{CN}}. In \cite{Tanner_GLDPC}
\emph{regular} \ac{GLDPC} codes (also known as Tanner codes) were
investigated, these being \ac{GLDPC} codes where the \acp{VN} are
all repetition codes of the same length and the
\acp{CN} are all linear block codes of the same type.

\begin{figure*}[t]
\begin{center}
\psfrag{n2edges}{\small{$s$ edges}} \psfrag{k1bits}{ \small{$k$
bits}} \psfrag{n2k2genCN}{$(s,h)$ generalized CN}
\psfrag{n2k2eqns}{\small{$s-h$ equations}} \psfrag{SPCCN}{SPC CN}
\psfrag{repVN}{rep. VN} \psfrag{n1k1genVN}{$(q,k)$ generalized VN}
\psfrag{brace}{$\underbrace{\phantom{---------------------------}}$}
\psfrag{encodedbits}{$N$ code bits} \psfrag{n1edges}{\small{$q$
edges}}
\includegraphics[width=6 cm, angle=270]{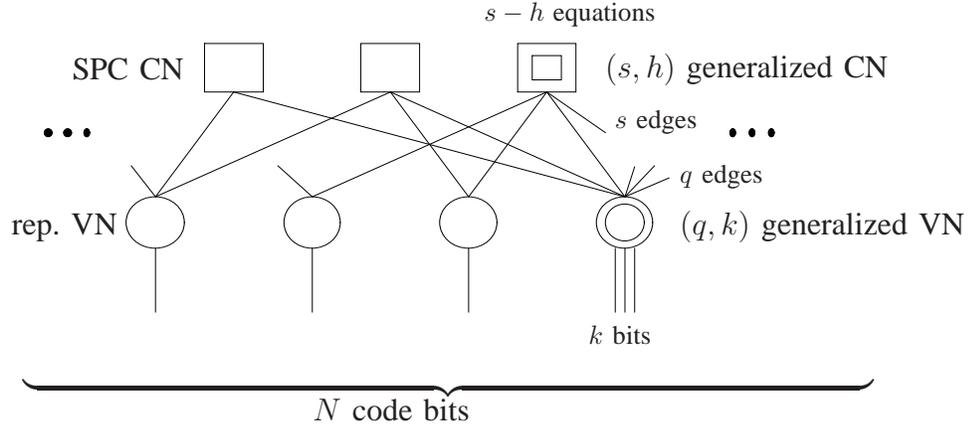}
\end{center}
\caption{Structure of a D-GLDPC code.} \label{fig:DGLDPC}
\end{figure*}

The growth rate of the weight distribution of binary \ac{GLDPC} codes was
investigated in
\cite{boutros99:generalized,lentmaier99:generalized,Tillich04:weight,paolini08:weight}.
In \cite{boutros99:generalized} the growth rate is calculated for
Tanner codes with BCH check component codes and length-2
repetition \acp{VN}, leading to an asymptotic lower bound on the
minimum distance. The same lower bound is developed in
\cite{lentmaier99:generalized} assuming Hamming \acp{CN} and length-2 repetition \acp{VN}. Both works extend
the approach developed by Gallager in \cite[Chapter 2]{Gallager}
to show that these ensembles have good growth rate behavior. The growth rate of the number of small
weight codewords for \ac{GLDPC} codes with a uniform \ac{CN} set (all \ac{CN} of the same type) and an irregular \ac{VN} set
(repetition \acp{VN} with different lengths) is investigated in \cite{Tillich04:weight}. It
is shown that the ensemble has good growth rate behavior when either the uniform \ac{CN} set
is composed of linear block codes with minimum distance at least $3$, or the
minimum length of the repetition \acp{VN} is 3. On the other hand,
if the minimum distance of the \acp{CN} and the minimum length of the
repetition \acp{VN} are both equal to 2, the goodness or otherwise of the growth rate behavior of the ensemble depends on the sign of the first order coefficient in
the growth rate Taylor series expansion. The results developed in
\cite{Tillich04:weight} were further extended in
\cite{paolini08:weight} to \ac{GLDPC} ensembles with an irregular
\ac{CN} set (\acp{CN} of different types). It was there
proved that, provided that there exist \acp{CN} with minimum distance $2$, a
parameter $\lambda'(0)C$, generalizing the parameter
$\lambda'(0)\rho'(1)$ of \ac{LDPC} code ensembles, plays in the context
of the weight distribution of \ac{GLDPC} codes the same role played by
$\lambda'(0)\rho'(1)$ in the context of the weight
distribution of \ac{LDPC} codes. The parameter $C$ is defined in
Section~\ref{section:further_def_&_notation}.

Interestingly, this latter results extends to binary \ac{GLDPC} codes the
same connection between the statistical properties of the weight distribution of irregular
codes and the stability condition over the
\ac{BEC}. In fact, it was shown in \cite{paolini08:stability} that
the stability condition of binary irregular \ac{GLDPC} codes over the
\ac{BEC} is given by
\begin{align}\label{eq:stability_GLDPC}
q^* \leq \left[ \lambda'(0)C \right]^{-1}.
\end{align}

Generalized \ac{LDPC} codes represent a promising solution for
low-rate channel coding schemes, due to an overall rate loss
introduced by the generalized \acp{CN}
\cite{miladinovic08:generalized}. Doubly-generalized LDPC (D-GLDPC) codes
generalize the concept of \ac{GLDPC} codes while facilitating much greater design 
flexibility in terms of code rate \cite{Wang_Fossorier_DG_LDPC} (an analogous idea may be
found in the previous work \cite{Dolinar}). In a D-GLDPC code,
the \acp{VN} as well as the \acp{CN} may be of any generic linear block code
types. A degree-$q$ \ac{VN} may in principle be any $(q,k)$ linear
block code, $q$ being the code length and $k$ the code dimension.
Such a \ac{VN} is associated with $k$ D-GLDPC code bits. It interprets these bits as its local information bits and interfaces
to the \ac{CN} set through its $q$ local code bits. A \ac{VN} which corresponds to a 
linear block code which is not a repetition code is said to be a
\emph{generalized \ac{VN}}. A D-GLDPC code is
said to be \emph{regular} if all of its \acp{VN} are of the same type
and all of its \acp{CN} are of the same type and is said to be \emph{irregular}
otherwise\footnote{Note that \acp{VN} associated with different representations of the same linear block code (i.e. with different generator matrices) are regarded as belonging to different types.}. The structure of a D-GLDPC code is depicted in
Fig.~\ref{fig:DGLDPC}.

A related class of bipartite-graph codes was considered in \cite{Barg_Zemor_expander_codes} where both \acp{CN} and \acp{VN} were generalized, but where the code bits were associated directly with the \emph{edges} of the Tanner graph (and thus the generator matrices associated with \acp{VN} were irrelevant). In this work it was shown that in certain regular code ensembles with the same local code of minimum distance $\ge 3$ at every \ac{CN} and \ac{VN}, asymptotically good codes exist in the ensemble which meet the Gilbert-Varshamov bound. These ensembles are generalizations of \emph{expander code} ensembles \cite{Sipser_Spielman_expander_codes}. Also, \cite{Barg_Mazumdar_Zemor_hypergraphs} presented similar results in the context of regular \emph{hypergraph} codes with random coding at the nodes (for a fixed hypergraph), random hypergraphs (with the same local code at every \ac{CN} and \ac{VN}), and random selection of both hypergraph and local codes. Also, \cite{Jason:JSAC} investigates the  asymptotic weight enumerators of many \ac{LDPC}-like codes including turbo codes and repeat-accumulate codes. 

In this paper the growth rate of the weight distribution of binary irregular D-GLDPC codes
is analyzed for small weight codewords. It is shown that a given irregular D-GLDPC code ensemble has good growth rate behavior when there are no \acp{VN} with minimum distance $2$, and likewise when there are no \acp{CN} with minimum distance $2$. It is also shown that, in the case where there exist both \acp{VN}
and \acp{CN} with minimum distance $2$, a parameter $1/P^{-1}(1/C)$
discriminates between an asymptotically small and an
asymptotically large expected number of small linear-weight
codewords (the function $P(x)$ is defined in
Section~\ref{section:further_def_&_notation}). The parameter
$1/P^{-1}(1/C)$ generalizes the above mentioned parameters
$\lambda'(0)\rho'(1)$ and $\lambda'(0)C$ to the case where both
generalized \acp{VN} and generalized \acp{CN} are present. The
obtained result also represents the extension to the D-GLDPC
case of the previously recalled connection with the stability
condition over the \ac{BEC}. In fact, it was proved in
\cite[Theorem 2]{paolini08:stability} that the stability condition of
D-GLDPC codes over the \ac{BEC} is given by
\begin{align}\label{eq:stability_D-GLDPC}
q^* \leq P^{-1}(1/C)\, .
\end{align}

The paper is organized as follows. Section~\ref{section:irregular_D_GLDPC} defines the D-GLDPC ensemble of interest, and introduces some definitions and notation pertaining to this ensemble. Section~\ref{section:further_def_&_notation} defines further terms regarding the \acp{VN} and \acp{CN} which compose the D-GLDPC codes in the ensemble. Section~\ref{section:growth_rate} presents the main result of the paper regarding the growth rate of the weight distribution, together with several corollaries. Section~\ref{section:proof_of_main_result} proves this main result, and Section~\ref{section:conclusion} concludes the paper.

\section{Irregular Doubly-Generalized LDPC Code Ensemble}
\label{section:irregular_D_GLDPC}

We define a D-GLDPC code ensemble $\cM_n$ as follows, where $n$ denotes the number of \acp{VN}. There are $n_c$ different \ac{CN} types $t \in I_c = \{ 1,2,\cdots, n_c\}$, and $n_v$ different \ac{VN} types $t \in I_v = \{ 1,2,\cdots, n_v\}$. For each \ac{CN} type $t \in I_c$, we denote by $h_t$, $s_t$ and $r_t$ the \ac{CN} dimension, length and minimum distance, respectively. For each \ac{VN} type $t \in I_v$, we denote by $k_t$, $q_t$ and $p_t$ the \ac{VN} dimension, length and minimum distance, respectively. For $t \in I_c$, $\rho_t$ denotes the fraction of edges connected to \acp{CN} of type $t$. Similarly, for $t \in I_v$, $\lambda_t$ denotes the fraction of edges connected to \acp{VN} of type $t$. Note that all of these variables are independent of $n$.

The polynomials $\rho(x)$ and $\lambda(x)$ are defined by
\[
\rho(x) = \sum_{t\in I_c} \rho_t x^{s_t - 1}
\]   
and
\[
\lambda(x) = \sum_{t \in I_v} \lambda_t x^{q_t - 1} \; .
\]   
If $E$ denotes the number of edges in the Tanner graph, the number of \acp{CN} of type $t\in I_c$ is then given by $E \rho_t / s_t$, and the number of \acp{VN} of type $t\in I_v$ is then given by $E \lambda_t / q_t$. Denoting as usual $\int_0^1 \rho(x) \, {\rm d} x$ and $\int_0^1 \lambda(x) \, {\rm d} x$ by $\int \rho$ and $\int \lambda$ respectively, we see that the number of edges in the Tanner graph is given by
\[
E = \frac{n}{\int \lambda}
\]
and the number of \acp{CN} is given by $m = E \int \rho$. Therefore, the fraction of \acp{CN} of type $t \in I_c$ is given by
\begin{equation}
\gamma_t = \frac{\rho_t}{s_t \int \rho}
\label{eq:gamma_t_definition}
\end{equation}
and the fraction of \acp{VN} of type $t \in I_v$ is given by
\begin{equation}
\delta_t = \frac{\lambda_t}{q_t \int \lambda} \; .
\label{eq:delta_t_definition}
\end{equation}
Also the length of any D-GLDPC codeword in the ensemble is given by 
\begin{equation}
N = \sum_{t \in I_v} \left( \frac{E \lambda_t}{q_t} \right) k_t = \frac{n}{\int \lambda} \sum_{t \in I_v} \frac{\lambda_t k_t}{q_t} \; .
\label{eq:DG_LDPC_codeword_length}
\end{equation}
Note that this is a linear function of $n$. Similarly, the total number of parity-check equations for any D-GLDPC code in the ensemble is given by
\[
M = \frac{m}{\int \rho} \sum_{t \in I_c} \frac{\rho_t (s_t-h_t)}{s_t} \; .
\]
A code in the irregular D-GLDPC ensemble then corresponds to a permutation on the $E$ edges connecting \acp{CN} to \acp{VN}. The \emph{design rate} of the D-GLDPC ensemble is given by
\begin{equation}
R = 1 - \frac{M}{N} = 1 - \frac{\sum_{t \in I_c} \rho_t (1 - R_t)}{\sum_{t \in I_v} \lambda_t R_t}
\label{eq:design_rate}
\end{equation}
where for $t \in I_c$ (resp. $t \in I_v$), $R_t$ is the local code rate of \acp{CN} (resp. \acp{VN})
of type $t$. Each code in the ensemble has a code rate larger than or equal to $R$.

The growth rate of the weight distribution of the irregular D-GLDPC ensemble sequence $\{ \cM_n \}$ is defined by 
\begin{equation}
G(\alpha) = \lim_{n\rightarrow \infty} \frac{1}{n} \log \mathbb{E}_{\cM_n} \left[ N_{\alpha n} \right]
\label{eq:growth_rate_result}
\end{equation}
where $\mathbb{E}_{\cM_n}$ denotes the expectation operator over the ensemble $\cM_n$, and $N_{w}$ denotes the number of codewords of weight $w$ of a randomly chosen D-GLDPC code in the ensemble $\cM_n$. The limit in (\ref{eq:growth_rate_result}) assumes the inclusion of only those positive integers $n$ for which $\alpha n \in \mathbb{Z}$ and $\mathbb{E}_{\cM_n} [ N_{\alpha n} ]$ is positive (i.e., where the expression whose limit we seek is well defined). Note that the argument of the growth rate function $G(\alpha)$ is equal to the ratio of D-GLDPC codeword length to the number of \acp{VN}; by (\ref{eq:DG_LDPC_codeword_length}), this captures the behaviour of codewords linear in the block length, as in \cite{Di_Richardson_Urbanke} for the \ac{LDPC} case. 
\medskip
\begin{definition}
Let $G(\alpha)$ be the growth rate of the weight distribution of an irregular D-GLDPC ensemble
sequence. The \emph{critical exponent codeword weight ratio} is defined as $\alpha^*=\inf \{ \alpha
> 0 \; | \; G(\alpha)\geq 0 \}$. Also, the ensemble sequence is said to have \emph{good growth rate
behavior} if $\alpha^* > 0$, and \emph{bad growth rate behavior} if $\alpha^* = 0$.
\label{def:critical_exponent}
\end{definition}
\medskip
Thus an irregular D-GLDPC code ensemble sequence has good growth rate behavior if and only if it
contains an asymptotically small expected number of small linear-weight codewords. Note that an ensemble with good growth
rate behavior must necessarily contain asymptotically good code sequences. The present definition of
the critical exponent codeword weight ratio may also be found in \cite{orlitsky05:stopping}.

We next define the concepts of \emph{assignment} and \emph{split assignment}. The concept of
\emph{assignment} was used in \cite{Di_Richardson_Urbanke} to develop an expression for the growth
rate of the weight distribution of irregular \ac{LDPC} code ensembles. The concept
of split assignment is introduced in this paper.
\medskip
\begin{definition}
An \emph{assignment} is a subset of the edges of the Tanner graph. An assignment is said to have \emph{weight} $k$ if it has $k$ elements. An assignment is said to be \emph{check-valid} if the following condition holds: supposing that each edge of the assignment carries a $1$ and each of the other edges carries a $0$, each \ac{CN} recognizes a valid local codeword. 
\end{definition}
\medskip
\begin{definition}
A \emph{split assignment} is an assignment, together with a subset of the D-GLDPC code bits (called a \emph{codeword assignment}). A split assignment is said to have \emph{split weight} $(u, v)$ if its assignment has weight $v$ and its codeword assignment has $u$ elements. A split assignment is said to be \emph{check-valid} if its assignment is check-valid. A split assignment is said to be \emph{variable-valid} if the following condition holds: supposing that each edge of its assignment carries a $1$ and each of the other edges carries a $0$, and supposing that each D-GLDPC code bit in the codeword assigment is set to $1$ and each of the other code bits is set to $0$, each \ac{VN} recognizes a local input word and the corresponding valid local codeword.     
\end{definition}
\medskip
Note that for any D-GLDPC code, there is a bijective correspondence between the set of D-GLDPC codewords and the set of split assignments which are both variable-valid and check-valid. 
\section{Further Definitions and Notation}
\label{section:further_def_&_notation}
The weight enumerating polynomial for \ac{CN} type $t \in I_c$ is given by 
\begin{eqnarray*}
A^{(t)}(x) & = & \sum_{u=0}^{s_t} A_u^{(t)} x^u \\
& = & 1 + \sum_{u=r_t}^{s_t} A_u^{(t)} x^u \; .
\end{eqnarray*}
Here $A_u^{(t)} \ge 0$ denotes the number of weight-$u$ codewords for \acp{CN} of type $t$. Note that $A_{r_t}^{(t)} > 0$ for all $t \in I_c$. Also, for each $t \in I_c$, corresponding to the polynomial $A^{(t)}(x)$ we denote the sets
\begin{equation}
U_t = \{ i \in \mathbb{N} \; : \; A^{(t)}_{i} > 0 \}
\label{eq:Ut}
\end{equation}
and
\begin{equation}
U_t^{-} = U_t \backslash \{ 0 \} \; .
\label{eq:Ut-}
\end{equation}

The bivariate weight enumerating polynomial for \ac{VN} type $t \in I_v$ is given by 
\begin{eqnarray*}
B^{(t)}(x,y) & = & \sum_{u=0}^{k_t} \sum_{v=0}^{q_t} B_{u,v}^{(t)} x^u y^v \\
& = & 1 + \sum_{u=1}^{k_t} \sum_{v=p_t}^{q_t} B_{u,v}^{(t)} x^u y^v \; .
\end{eqnarray*}
Here $B_{u,v}^{(t)} \ge 0$ denotes the number of weight-$v$ codewords generated by input words of weight $u$, for \acp{VN} of type $t$. Also, for each $t \in I_v$, corresponding to the polynomial $B^{(t)}(x,y)$ we denote the sets
\begin{equation}
S_t = \{ (i,j) \in \mathbb{N}^2 \; : \; B^{(t)}_{i,j} > 0 \}
\label{eq:St}
\end{equation}
and
\begin{equation}
S_t^{-} = S_t \backslash \{ (0,0) \} \; .
\label{eq:St-}
\end{equation}
We also define
\begin{equation}
S^{-} = \cup_{t \in I_v} S_t^{-} \; .
\end{equation}

We denote the smallest minimum distance over all \ac{CN} types by
\[
r = \min \{ r_t \; : \; t \in I_c \} \ge 2
\]
and the set of \ac{CN} types with this minimum distance by
\[
X_c = \{ t \in I_c \; : \; r_t = r \} \; .
\]
We define the parameter 
\begin{equation}
\psi = r/(r-1)
\label{eq:psi}
\end{equation}
and note that we have $1 < \psi \le 2$ with equality if and only if $r = 2$. We define the parameter
\begin{equation}
C = r \sum_{t \in X_c} \frac{\rho_t A^{(t)}_{r}}{s_t} > 0 \; .
\label{eq:C_definition}
\end{equation}
We also define $\bar{r}$ as the smallest integer $i > r$ such that there exists some CN with a non-zero number of weight-$i$ codewords:
\begin{equation}
\bar{r} = \min\{i > r \; : \; A_i^{(t)} > 0 \textrm{ for some } t \in I_c \textrm{ and } i \in U_t^-\} \, .
\label{eq:rbar_definition}
\end{equation}
The parameter $\bar{r}$ represents the second smallest minimum distance over all \ac{CN} types.
  
Similarly, we denote the smallest minimum distance over all \ac{VN} types by
\[
p = \min \{ p_t \; : \; t \in I_v \} \ge 2
\]
and the set of \ac{VN} types with this minimum distance by 
\[
X_v = \{ t \in I_v \; : \; p_t = p \} \; .
\]
We also define $\bar{p}$ as the smallest integer $j > p$ such that there exists some VN with a non-zero number of weight-$j$ codewords:
\begin{equation}
\bar{p} = \min\{j > p \; : \; B_{i,j}^{(t)} > 0 \textrm{ for some } t \in I_v \textrm{ and } (i,j) \in S_t^-\} \, .
\label{eq:pbar_definition}
\end{equation}
The parameter $\bar{p}$ represents the second smallest minimum distance over all \ac{VN} types.

For each $(i,j) \in S^{-}$, define
\begin{equation}
T_{i,j} = \frac{j-\psi}{i} \; ,
\end{equation}
and define the parameter
\begin{equation}
T = \min_{ (i,j) \in S^{-} } \left\{ T_{i,j} \right\}
\label{eq:T}
\end{equation}
and the set
\[
Y_v = \left\{ t \in I_v \; : \; \min_{ (i,j) \in S_t^{-} } \left\{ T_{i,j} \right\} = T \right\} \; .
\]
We also define the parameter 
\begin{equation}
\chi = \min_{(i,j) \in S^{-} : T_{i,j} > T} \left\{ (T_{i,j} - T) i \right\} \, .
\label{eq:chi_definition}
\end{equation}
Since $1 < \psi \le 2$ with equality if and only if $r=2$, and $j \ge p \ge 2$ for all $(i,j) \in S^{-}$, it follows that $T \ge 0$ with equality if and only if $r=p=2$. Also, for $t \in Y_v$, define
\begin{equation}
P_t = \left\{ (i,j) \in S_t^{-} \; : \; \frac{j-\psi}{i} = T \right\} \; .
\label{eq:P_t_definition}
\end{equation}
Note that in the specific case $r= p = 2$, we have $T = 0$ and $Y_v=X_v$, and we may write $P_t = \{ (i,2) \; : \; i \in L_t \}$ where $L_t = \{ i \in \mathbb{N} \; : \; B^{(t)}_{i,2} > 0 \}$ for each $t \in X_v$ -- note that these sets are nonempty. 

We define the polynomials
\begin{equation}
Q_1(x) = \sum_{t \in Y_v} \frac{\lambda_t}{q_t} \sum_{ (i,j) \in P_t } j B^{(t)}_{i,j} C^{j/r} \left( \frac{\int \lambda}{e} \right)^{iT/\psi} x^i
\label{eq:P1x_definition}
\end{equation}
and
\begin{equation}
Q_2(x) = \sum_{t \in Y_v} \frac{\lambda_t}{q_t} \sum_{ (i,j) \in P_t } i B^{(t)}_{i,j} C^{j/r} \left( \frac{\int \lambda}{e} \right)^{iT/\psi} x^i \; .
\label{eq:P2x_definition}
\end{equation}
Since all of the coefficients of $Q_1(x)$ and $Q_2(x)$ are positive, these polynomials are both monotonically increasing on $[0,\infty)$ and therefore their inverses, denoted by $Q_1^{-1}(x)$ and $Q_2^{-1}(x)$ respectively, are well-defined and unique on this interval. Note that in the case $r=p=2$, we have 
\[
Q_1(x) = C \cdot P(x)
\]
where
\begin{equation}
P(x) = 2 \sum_{t \in X_v} \frac{\lambda_t}{q_t} \sum_{i \in L_t} B^{(t)}_{i,2} x^i \; .
\label{eq:Pofx_definition}
\end{equation}
Also note that in the case $r=p=2$, (\ref{eq:C_definition}) becomes 
\begin{equation}
C = 2 \sum_{t \in X_c} \frac{\rho_t A^{(t)}_{2}}{s_t} 
\label{eq:C_definition_revisited}
\end{equation}
and we define
\begin{equation}
V = 2 \sum_{t \in X_v} \frac{\lambda_t B^{(t)}_{2}}{q_t} > 0
\label{eq:V_definition}
\end{equation}
as the counterpart of the parameter $C$ in the variable node domain. Here $B^{(t)}_{2} = \sum_{i \in L_t} B^{(t)}_{i,2}$ is the total number of weight-$2$ codewords for \acp{VN} of type $t$.
Note that in this case the parameter $C$ depends only on the \acp{CN} with minimum distance $2$, and the parameter $V$ and the polynomial $P(x)$ depend only on the \acp{VN} with minimum distance $2$.
Also note that while the polynomial $P(x)$ given by (\ref{eq:Pofx_definition}) depends on the \ac{VN} \emph{representations} (i.e. generator matrices), the parameter $V$ given by (\ref{eq:V_definition}) does not.

Throughout this paper, we make use of the following standard notation. Let $g(x)$ be a nonnegative real-valued function, and let $f(x)$ be a real-valued function. We say that $f(x)$ is $O \left( g(x) \right)$, writing $f(x) \sim O \left( g(x) \right)$, if and only if there exist positive real numbers $k$ and $\epsilon$, both independent of $x$, such that 
\[
\left| f(x) \right| \le k g(x) \quad \forall \; 0 \le x \le \epsilon \; .
\]
Let $a(n)$ and $b(n)$ be two real-valued sequences, where $b(n) \ne 0$ for all $n$, and let $q(n) = a(n)/b(n)$. We say that $a(n)$ is \emph{asymptotically equal} to $b(n)$ as $n \rightarrow \infty$, writing $a(n) \asympequaln b(n)$, if and only if $\lim_{n\rightarrow \infty} q(n) = 1$.

Finally, throughout this paper, the notation $e = \exp(1)$ denotes Napier's number. 

\section{Growth Rate for Doubly-Generalized \ac{LDPC} Code Ensemble}
\label{section:growth_rate}
The following theorem constitutes the main result of the paper.
\medskip
\begin{theorem}
Consider an irregular D-GLDPC code ensemble sequence $\cM_n$. The growth rate of the weight distribution is given by
\begin{equation}
G(\alpha) = \frac{T}{\psi} \, \alpha \log \alpha + \alpha \Bigg[ \log \left( \frac{1}{Q_1^{-1}(1)} \right) + \frac{T}{\psi} \log \left( \frac{1}{Q_2(Q_1^{-1}(1))} \right) \Bigg] + O(\alpha^{\xi}) \; ,
\label{eq:growth_rate_general}
\end{equation}
where 
\begin{equation}
\xi = \min \left\{ \frac{\bar{r}}{r}, \frac{\chi}{\psi} + 1, 2 \right\} \; .
\label{eq:xi_definition}
\end{equation}
\label{thm:growth_rate} 
\end{theorem}

This theorem is proved in Section \ref{section:proof_of_main_result}. We next provide a series of corollaries to this result; this serves to illustrate the manner in which several related results in the literature follow as special cases of Theorem \ref{thm:growth_rate}.
\medskip
\begin{corollary}
In the case where either $r>2$ or $p>2$, the growth rate of the weight distribution is given by
\begin{equation}
G(\alpha) = \frac{T}{\psi} \, \alpha \log \alpha + O(\alpha) \; ,
\label{eq:growth_rate_case_2}
\end{equation}
where $T > 0$.  
\label{cor:1}
\end{corollary}
\medskip
Thus if either $r>2$ or $p>2$, we have $\alpha^*>0$ and the ensemble sequence exhibits good growth rate behavior. This generalizes results along this line in \cite{Tillich04:weight,paolini08:weight,Barg_Zemor_expander_codes}. A special case of Corollary \ref{cor:1} is as follows.
\medskip
\begin{corollary}
Suppose $r>2$ or $p>2$ and also $\cup_{t \in Y_v} P_t = \{ (i,j) \}$ for a single point $(i,j)$, i.e. a single point $(i,j)$ achieves the minimum in (\ref{eq:T}) although this $(i,j)$ may be manifest in different \ac{VN} types $t \in Y_v$. Then
\begin{equation}
G(\alpha) = \frac{T}{\psi} \, \alpha \log \alpha + K \alpha + O(\alpha^{\xi}) \; ,
\label{eq:growth_rate_case_3}
\end{equation} 
where $K$ is given by
\begin{equation}
K = \frac{1}{i} \Bigg[ \log \left( i \sum_{t \in Y_v} B^{(t)}_{i,j} \delta_t \right) + \frac{j}{r} \log C + \frac{j}{\psi} \log \left( \frac{j \int \lambda}{i} \right) \Bigg] - \frac{T}{\psi} \; .
\label{eq:K}
\end{equation}
\label{cor:2}
\end{corollary}
\medskip
\begin{proof} 
This result follows by making the appropriate substitutions in Theorem \ref{thm:growth_rate} and noting that in this case $Q_1(x)$ and $Q_2(x)$ are monomials satisfying $Q_2(Q_1^{-1}(1)) = i/j$. 
\end{proof}
\medskip
\begin{corollary}
Consider a \ac{GLDPC} code ensemble with irregular \ac{CN} set and irregular \ac{VN} set (i.e. different \ac{VN} degrees). Let $r$ denote the smallest minimum distance of the \acp{CN}, and $p$ denote the minimum \ac{VN} degree. Then
\begin{equation}
G(\alpha) = \left( p - \frac{p}{r} - 1 \right) \alpha \log \alpha + K \alpha + O(\alpha^{\xi}) \; ,
\label{eq:growth_rate_GLDPC_general}
\end{equation} 
where
\begin{equation}
K = \log \left( e \tilde{\delta} \right) + \frac{p}{r} \log C + \frac{p}{\psi} \log \left( \frac{p \int \lambda}{e} \right) \; .
\label{eq:K_GLDPC}
\end{equation}
where $\tilde{\delta}$ represents the fraction of VNs of degree $p$. 
\label{cor:3}
\end{corollary}
\medskip
\begin{proof} 
In this case, each VN type $t \in I_v$ satisfies $B^{(t)}(x,y) = 1 + x y ^{p_t}$. Let $\tilde{t} \in I_v$ represent the VN type with minimum length (degree), i.e., $p_{\tilde{t}} = p$, and note that $\tilde{\delta} = \delta_{\tilde{t}}$; then $Y_v = \left\{ \tilde{t} \right\}$ and $\cup_{t \in Y_v} P_t = P_{\tilde{t}} = \{ (1,p) \}$. Application of Corollary \ref{cor:2} then directly yields the required result, where we use the fact that $T = p - \psi$ in this case.
\end{proof}
\medskip
This provides a generalization of the result of \cite{Tillich04:weight} which derived (\ref{eq:growth_rate_GLDPC_general}) for the case of \ac{GLDPC} codes with \emph{regular} \ac{CN} sets and irregular \ac{VN} degrees, and which did not include the result (\ref{eq:K_GLDPC}) regarding the evaluation of the parameter $K$.
\medskip
\begin{corollary}
Consider a D-GLPDC code ensemble $\cM_n$ satisfying $r=p=2$. Then the growth rate of the weight distribution is given by
\begin{equation}
G(\alpha) = \alpha \log \left[ \frac{1}{P^{-1}(1/C)} \right] + O(\alpha^{\xi}) \; ,
\label{eq:growth_rate_case_1}
\end{equation}
where the polynomial $P(x)$ and the parameter $C$ are given by (\ref{eq:Pofx_definition}) and (\ref{eq:C_definition_revisited}) respectively, and where
\begin{equation}
\xi = \min \left\{ \frac{\bar{r}}{2}, \frac{\bar{p}}{2}, 2 \right\} = \left\{ \begin{array}{cc}
2 & \textrm{ if } \bar{r} > 3 \textrm{ or } \bar{p} > 3 \\
3/2 & \textrm{ otherwise. }\end{array}\right. \;
\label{eq:symmetric_exponent_rp2}
\end{equation} 
\label{cor:4}
\end{corollary}
\medskip
\begin{proof} 
When $r=p=2$, we have $T=0$ and $\psi = 2$; also by (\ref{eq:chi_definition}) we have
\[
\chi = \min_{(i,j) \in S^{-} : T_{i,j} > 0} \{ i T_{i,j} \} = \min_{(i,j) \in S^{-} : j > 2} \{ j-\psi \} = \bar{p}-2 \; ,
\]
which implies that $\chi/\psi + 1 = \bar{p}/2$. Also, it may be verified that $Q_1^{-1}(1) = P^{-1}(1/C)$ in this case.
\end{proof}
\medskip
Corollary \ref{cor:4} first appeared in \cite{flanagan08:growth}. A necessary and sufficient condition for a D-GLDPC ensemble satisfying $r=p=2$ to have good growth rate behavior follows in a straightforward manner as shown next. 
\medskip
\begin{corollary}
Consider a D-GLPDC code ensemble $\cM_n$ satisfying $r=p=2$. Then, a necessary and sufficient condition for $\cM_n$ to have good growth rate behavior is 
\begin{equation}
C \cdot V < 1 
\end{equation}
where $C$ and $V$ are given by (\ref{eq:C_definition_revisited}) and (\ref{eq:V_definition}) respectively.
\label{cor:5}
\end{corollary}
\medskip
\begin{proof}
From (\ref{eq:growth_rate_case_1}) the necessary and sufficient condition is $1/P^{-1}(1/C) < 1$; rearranging and using the monotonicity of $P(x)$ yields the result.
\end{proof}
\medskip
It is worth noting that an analogous theorem to Theorem \ref{thm:growth_rate} holds also for the stopping set size distribution of irregular D-GLDPC codes. The proof is almost identical to that of Theorem \ref{thm:growth_rate} and is outlined in Appendix \ref{app:stopping_sets}.

\subsection{Discussion}\label{subsection:graphical_interpretation}

\begin{figure}[t]
\begin{center}
\psfrag{y=tx+psi}[lb]{\small{$\!\!\!\mathcal{L}:j=\frac{1}{2}i
+\frac { 3 } { 2 } $ } } \psfrag{i}{
\small{$i$}} \psfrag{jminuspsioveri}[lt]{$\,\,\,\,\,\frac{j-\psi}{i}$}
\psfrag{increasing}[lt]{\small{increasing}}
\psfrag{j}{$j$} \psfrag{psi}{\small{$\psi$}} \psfrag{fp}[c]{\small{fixed point$\phantom{---...}$}}
\psfrag{0ph}[tc]{\small{$(0,\psi)\phantom{.....}$}} \psfrag{ar}[c]{\Large{$\rightarrow$}}
\includegraphics[width=13 cm, angle=0]{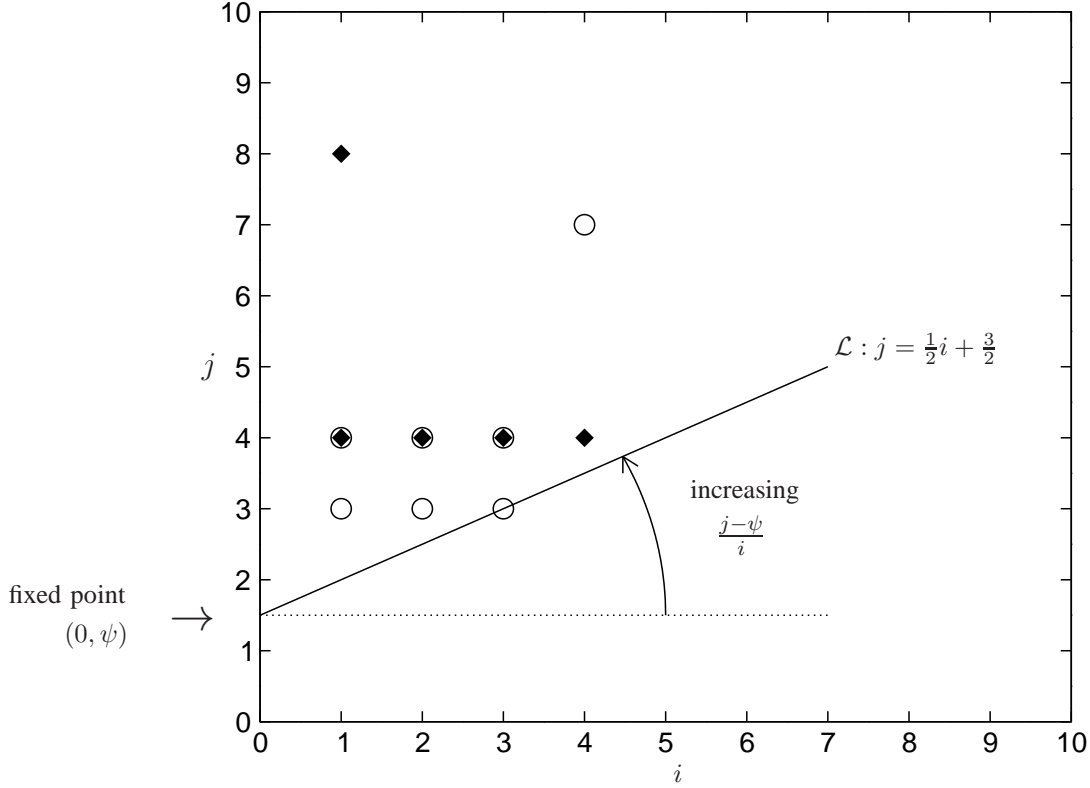}
\end{center}
\caption{Diagram of the VN input-output weight enumerating functions and growth rate dominant set
in the $(i,j)$ plane. The illustration is for a smallest CN minimum distance of $r=3$, so $\psi =
3/2$. The sets $S_t^-$ for the two VN types $t \in I_v = \{1,2\}$ are illustrated by open circles
and filled diamonds respectively. The line $\mathcal{L}$ has a fixed point at $(0,\psi)$ and is
rotated in an anticlockwise fashion until it touches any of these points. In this example, this
occurs at the point $(3,3) \in S_1^-$. Therefore in this example $T = 1/2$, $Y_v = \{ 1 \}$,
$P_1 = \{ (3,3) \}$ and for small values of $\alpha$ the dominating contribution to the growth rate
comes from weight-$3$ local codewords of the type-$1$ VNs generated by weight-$3$ local input
words.}\label{fig:geometric_construction}
\end{figure}

From Theorem~\ref{thm:growth_rate} and from the definitions of $Q_1(x)$ and $Q_2(x)$ given in
\eqref{eq:P1x_definition} and \eqref{eq:P2x_definition} respectively, we observe
that the triples $(t,i,j)$ (or, equivalently, VN input-output weight enumerating function
coefficients $B^{(t)}_{i,j}$) such that $(i,j)$ lies in one of the sets $P_t$ ($t \in Y_v$) make a
dominating contribution to the growth rate for values of $\alpha$ close to zero. We will refer to
the set of such triples as the \emph{dominant set}. Note that the dominant set may equivalently be described as 
the set of triples $(t,i,j)$ such that $T_{i,j} = T$ for some $t \in I_v$, $(i,j) \in S_t^{-}$.

Interestingly, the dominant set admits an instructive graphical interpretation and may be easily
identified using a very simple geometric construction. In the $(i,j)$ plane, a line $\mathcal{L}$
through the fixed point $(0,\psi)$ is rotated in an anticlockwise fashion until it comes in contact
with one or more of the points $(i,j) \in S^-$. The slope of the line
$\mathcal{L}$ at this point is the parameter $T$ defined in~\eqref{eq:T}, the set of $t \in I_v$
which have points $(i,j) \in \mathcal{L}$ is the set defined as $Y_v$, and for each such $t$ the set
of such points on $\mathcal{L}$ is the set $P_t$ defined in~\eqref{eq:P_t_definition}. Note that due
to~\eqref{eq:psi}, the position of the fixed point $(0,\psi)$ depends only on the
smallest CN minimum distance $r$, and always lies somewhere on the line segment joining $(0,1)$ and
$(0,2)$, including the latter endpoint.

This interpretation is illustrated in Fig.~\ref{fig:geometric_construction} for an example D-GLDPC
code with two VN types $I_v = \{1,2\}$. Type $1 \in I_v$ (open
circles) corresponds to the Hamming code of length $q_1=7$, dimension $k_1=4$, minimum distance
$p_1=3$ and represented by the systematic generator matrix
$\mathbf{G}_{\mathsf{H}}=[1000111,0100110,0010101,0001011]$. Type $2 \in I_v$ (filled diamonds) corresponds to the first-order Reed-Muller code of length
$q_2=8$, dimension $k_2=4$, minimum distance $p_2=4$ and represented by the
generator matrix $\mathbf{G}_{\mathsf{RM}}=[11111111,11110000,11001100,10101010]$. In the
specific case of Fig.~\ref{fig:geometric_construction} we have $\psi=3/2$, which corresponds to
$r=3$.\footnote{For example, assuming a uniform CN set composed of $(15,11)$ Hamming codes, and
assuming $\lambda_1=7/15$ and $\lambda_2=8/15$, through \eqref{eq:design_rate} this would represent
a rate $R=1/2$ ensemble.} It may be deduced from the figure that for values of $\alpha$ close to zero the
growth rate $G(\alpha)$ is dominated by weight-$3$ local codewords of the
Hamming VNs associated with weight-$3$ local input words.

From Corollary \ref{cor:4} we observe that in the special case where $r=p=2$,  
the growth rate depends only on the \acp{CN} and \acp{VN} with
minimum distance equal to $2$, and (\ref{eq:growth_rate_case_1}) is a first-order Taylor series around $\alpha = 0$ which directly
generalizes the results of \cite{Di_Richardson_Urbanke} and \cite{paolini08:weight} (for irregular
\ac{LDPC} and \ac{GLDPC} codes respectively) to the case of irregular D-GLDPC codes. Note that the error term in \eqref{eq:growth_rate_case_1} is $O(\alpha^\xi)$, which is different to the error term $O(\alpha^2)$ reported in previous literature (Theorem 4.1 in \cite{flanagan08:growth}, Theorem 5 in \cite{Di_Richardson_Urbanke}); this difference is manifest in the case where $r=p=2$ and either $\bar{r}=3$ or $\bar{p}=3$. Corollary \ref{cor:4} indicates that in the analysis of the asymptotic growth rate of the weight
distribution, the parameter $1 / P^{-1}(1/C)$ in the context of D-GLDPC codes plays an analagous
role to the parameter $\lambda'(0) \rho'(1)$ for irregular \ac{LDPC} codes, and to the parameter
$\lambda'(0)C$ for irregular \ac{GLDPC} codes. It discriminates between ensemble
sequences with good growth rate behavior, for which $1/P^{-1}(1/C)<1$, and ensemble sequences
with bad growth rate behavior, for which $1/P^{-1}(1/C)>1$.

\section{Proof of the Main Result}
\label{section:proof_of_main_result}
In this section, Theorem \ref{thm:growth_rate} is proved. For ease of presentation, the proof is broken into four parts.
\subsection{Number of check-valid assignments of weight $\epsilon m$ over $\gamma m$ \acp{CN} of type $t \in I_c$}
Consider $\gamma m$ \acp{CN} of the same type $t \in I_c$. Using generating functions, the number of check-valid assignments (over these \acp{CN}) of weight $\epsilon m$ is given by\footnote{Here we make use of the following general result \cite{Wilf}. Let $a_i$ be the number of ways of obtaining an outcome $i\in\mathbb{Z}$ in experiment $\cA$, and let $b_j$ be the number of ways of obtaining an outcome $j\in\mathbb{Z}$ in experiment $\cB$. Also let $c_k$ be the number of ways of obtaining an outcome $(i,j)$ in the combined experiment $(\cA, \cB)$ with sum $i+j=k$. Then the generating functions $A(x)=\sum_i a_i x^i$, $B(x)=\sum_j b_j x^j$ and $C(x)=\sum_k c_k x^k$ are related by $C(x) = A(x) B(x)$.} 
\[
N_{c,t}^{(\gamma m)}(\epsilon m) = \coeff \left[ \left( A^{(t)}(x) \right) ^{\gamma m}, x^{\epsilon m} \right]
\]
where $\coeff [ p(x), x^c ]$ denotes the coefficient of $x^c$ in the polynomial $p(x)$. We now use the following result, the proof of which appears in \cite[Appendix A]{Burshtein_Miller}:
\medskip
\begin{lemma}
Let $A(x) = 1 + \sum_{u=c}^{d} A_u x^u$, where $1 \le c \le d$, be a polynomial satisfying $A_c > 0$ and $A_u \ge 0$ for all $c < u \le d$. For a fixed positive rational number $\xi$, consider the set of positive integers $\ell$ such that $\xi \ell \in \mathbb{N}$ and $\coeff [ \left( A(x) \right) ^{\ell}, x^{\xi \ell}] > 0$. Then either this set is empty, or it has infinite cardinality; if $t$ is one such $\ell$, then so is $jt$ for every positive integer $j$. In the latter case, the following limit is well defined and exists:
\begin{equation}
\lim_{\ell\rightarrow \infty} \frac{1}{\ell} \log \coeff \left[ \left( A(x) \right) ^{\ell}, x^{\xi \ell} \right]
= \max_{\bldsmallbeta} \sum_{i \in U} \beta_i \log \left( \frac{A_i}{\beta_i} \right)
\end{equation}
where $U = \{ i \in \mathbb{N} \; : \; A_i > 0 \}$, $\bldbeta = ( \beta_i )_{i \in U}$, and the maximization is subject to the constraints $\sum_{i \in U} \beta_i = 1$, $\sum_{i \in U} i \beta_i = \xi$ and $\beta_i \ge 0$ for all $i \in U$.
\end{lemma} 
\medskip
Applying this lemma by substituting $A(x) = A^{(t)}(x)$, $\ell=\gamma m$ and $\xi = \epsilon/\gamma$, we obtain that with $\gamma$ fixed, as $m \rightarrow \infty$
\begin{eqnarray}
N_{c,t}^{(\gamma m)}(\epsilon m) = \coeff \left[ \left( A^{(t)}(x) \right) ^{\gamma m}, x^{\epsilon m} \right] 
\label{eq:Nct_epsilon_start} \\
\asympequalm \exp \left\{ m \gamma \max_{\bldsmallbeta^{(t)}} \sum_{i \in U_t} \beta^{(t)}_{i} \log \left( \frac{A^{(t)}_{i}}{\beta^{(t)}_{i}} \right) \right\} 
\label{eq:Nct_epsilon_mid} \\
\triangleq \exp \left\{ m W^{(\gamma)}_t(\epsilon) \right\}
\label{eq:Nct_epsilon_end}
\end{eqnarray}
where the maximization over $\bldbeta^{(t)} = ( \beta^{(t)}_{i} )_{i \in U_t}$ is subject to the constraints $\sum_{i \in U_t} \beta^{(t)}_{i} = 1$, $\sum_{i \in U_t^{-}} i \beta^{(t)}_{i} = \epsilon / \gamma$ and $\beta^{(t)}_{i} \ge 0$ for all $i \in U_t$ (recall that the sets $U_t$ and $U_t^{-}$ are given by (\ref{eq:Ut}) and (\ref{eq:Ut-})).

\subsection{Number of check-valid assignments of weight $\delta m$}
Next we derive an expression, valid asymptotically, for the number of check-valid assignments of weight $\delta m$. For each $t \in I_c$, let $\epsilon_t m$ denote the portion of the total weight $\delta m$ apportioned to \acp{CN} of type $t$. Then $\epsilon_t \ge 0$ for each $t \in I_c$, and $\sum_{t \in I_c} \epsilon_t = \delta$. Also denote $\bldepsilon = (\epsilon_1 \; \epsilon_2 \; \cdots \; \epsilon_{n_c})$. The number of check-valid assignments of weight $\delta m$ satisfying the constraint $\bldepsilon$ is obtained by multiplying the numbers of check-valid assignments of weight $\epsilon_t m$ over $\gamma_t m$ \acp{CN} of type $t$, for each $t \in I_c$,
\begin{equation}
N_c^{(\bldepsilon)}(\delta m) = \prod_{t \in I_c} N_{c,t}^{(\gamma_t m)}(\epsilon_t m) 
\label{eq:Nc_epsilon}
\end{equation}
where the fraction $\gamma_t$ of \acp{CN} of type $t \in I_c$ is given by (\ref{eq:gamma_t_definition}).

The number of check-valid assignments of weight $\delta m$, which we denote $N_c(\delta m)$, is equal to the sum of $N_c^{(\bldepsilon)}(\delta m)$ over all admissible vectors $\bldepsilon$; therefore, as $m\rightarrow \infty$
\begin{equation}
N_c(\delta m) \, \asympequalm \sum_{\bldsmallepsilon \; : \; \sum_{t \in I_c} \epsilon_t = \delta}
\exp \left\{ m \sum_{t \in I_c} W^{(\gamma_t)}_t(\epsilon_t) \right\}
\label{eq:sum_of_exp_check}
\end{equation}
where we have used (\ref{eq:Nct_epsilon_end}) and (\ref{eq:Nc_epsilon}). However, the asymptotic expression as $m\rightarrow \infty$ is dominated by the distribution $\bldepsilon$ which maximizes the argument of the exponential\footnote{Observe that as $m\rightarrow \infty$, $\sum_t \exp ( m Z_t ) \asympequalm \exp ( m \max_t \{Z_t\} )$}. Therefore as $m\rightarrow \infty$
\begin{equation}
N_c(\delta m) \asympequalm \exp \left\{ m W \right\} 
\label{eq:N_c_max} 
\end{equation}
where
\begin{equation}
W = \max_{\bldsmallepsilon} \sum_{t \in I_c} \gamma_t \max_{\bldsmallbeta^{(t)}} \sum_{i \in U_t} \beta^{(t)}_{i} \log \left( \frac{A^{(t)}_{i}}{\beta^{(t)}_{i}} \right) \; ,
\label{eq:W} 
\end{equation}
the maximization over $\bldepsilon$ is subject to the constraint
\begin{equation}
\sum_{t \in I_c} \epsilon_t = \delta \; ,
\label{eq:sum_epsilont_constraint}
\end{equation}
and for each $t \in I_c$ the maximization over $\bldbeta^{(t)} = ( \beta^{(t)}_{i} )_{i \in U_t}$ is subject to the constraints 
\begin{equation}
\sum_{i \in U_t} \beta^{(t)}_i = 1
\label{eq:betat_sum_constraint}
\end{equation}
\begin{equation}
\sum_{i \in U_t^{-}} i \beta^{(t)}_i = \epsilon_t/\gamma_t
\label{eq:sum_t_betat_constraint}
\end{equation}
and
\begin{equation}
\beta^{(t)}_i \ge 0 \quad \forall i \in U_t \; .
\label{eq:nonnegative_betat_constraint}
\end{equation}
Next, for each $t \in I_c$ we define
\[
F_t(\bldbeta^{(t)}) = \beta^{(t)}_{0} \log \left( \frac{1}{\beta^{(t)}_{0}} \right) - \sum_{i \in U_t^{-}} \beta^{(t)}_{i} \; .
\]
We then have the following lemma.
\medskip
\begin{lemma}
The expression $\sum_{t \in I_c} \gamma_t F_t(\bldbeta^{(t)})$ is $O(\delta^2)$ for any $\bldbeta^{(t)}$ satisfying the optimization constraints~(\ref{eq:sum_epsilont_constraint})--(\ref{eq:nonnegative_betat_constraint}).
\label{lemma:O2_term_check}
\end{lemma}
\medskip
A proof of this lemma is given in Appendix \ref{app:proof_lemma}. It follows from Lemma \ref{lemma:O2_term_check} that the expression $\sum_{t \in I_c} \gamma_t F_t(\bldbeta^{(t)})$ is $O(\delta^2)$ for the maximizing $\bldbeta^{(t)}$. Therefore 
\begin{eqnarray*}
W & = & \max_{\bldsmallepsilon} \sum_{t \in I_c} \gamma_t \max_{\bldsmallbeta^{(t)}} \left[ \sum_{i \in U_t^{-}} \beta^{(t)}_{i} \log \left( \frac{e A^{(t)}_{i}}{\beta^{(t)}_{i}} \right) + F_t(\bldbeta^{(t)}) \right] \\
& = & \max_{\bldsmallepsilon} \sum_{t \in I_c} \gamma_t \max_{\bldsmallbeta^{(t)}} \sum_{i \in U_t^{-}} \beta^{(t)}_{i} \log \left( \frac{e A^{(t)}_{i}}{\beta^{(t)}_{i}} \right) + O(\delta^2) 
\end{eqnarray*}
where the maximization over $\bldbeta^{(t)} = ( \beta^{(t)}_{i} )_{i \in U_t^{-}}$ (for each $t \in I_c$) is subject to the constraint~(\ref{eq:sum_t_betat_constraint}) together with $\beta^{(t)}_i \ge 0$ for all $i \in U_t^{-}$. In what follows, for convenience of presentation we shall temporarily omit the $O(\delta^2)$ term in the expression for $W$.

Next we make the substitution $\theta^{(t)}_{i} = \gamma_t \beta^{(t)}_{i}$ for all $t \in I_c$, $i \in U_t^{-}$. This yields
\begin{equation*}
W = \max_{\bldsmallepsilon} \sum_{t \in I_c} \max_{\bldsmalltheta^{(t)}} \sum_{i \in U_t^{-}} \theta^{(t)}_{i} \log \left( \frac{e A^{(t)}_{i} \gamma_t}{\theta^{(t)}_{i}} \right)
\end{equation*}
where the maximization over $\bldtheta^{(t)} = ( \theta^{(t)}_{i} )_{i \in U_t^{-}}$ (for each $t \in I_c$) is subject to the constraints $\sum_{i \in U_t^{-}} i \theta^{(t)}_{i} = \epsilon_t$ and $\theta^{(t)}_{i} \ge 0$ for all $i \in U_t^{-}$.
We observe that this maximization may be recast as 
\begin{equation*}
W = \max_{\bldsmalltheta} \sum_{t \in I_c} \sum_{i \in U_t^{-}} \theta^{(t)}_{i} \log \left( \frac{e A^{(t)}_{i} \gamma_t}{\theta^{(t)}_{i}} \right)
\end{equation*}
where by~(\ref{eq:sum_epsilont_constraint}) the maximization, which is now over $\bldtheta = (\theta^{(t)}_{i})_{t \in I_c, i \in U_t^{-}}$, is subject to the constraints 
\[
\sum_{t \in I_c} \sum_{i \in U_t^{-}} i \theta^{(t)}_{i} = \delta
\]
and $\theta^{(t)}_{i} \ge 0$ for all $t \in I_c$, $i \in U_t^{-}$.

Making the substitution $\upsilon^{(t)}_{i} = \theta^{(t)}_{i} / \delta$ for all $t \in I_c$, $i \in U_t^{-}$, we obtain
\begin{equation}
W = \delta \max_{\bldsmallupsilon} \sum_{t \in I_c} \sum_{i \in U_t^{-}} \upsilon^{(t)}_{i} \log \left( \frac{e A^{(t)}_{i} \gamma_t}{\delta \upsilon^{(t)}_{i}} \right)
\label{eq:W_defn_for_Lagrange}
\end{equation}
where the maximization over $\bldupsilon = (\upsilon^{(t)}_{i})_{t \in I_c, i \in U_t^{-}}$ is subject to the constraints 
\begin{equation}
\sum_{t \in I_c} \sum_{i \in U_t^{-}} i \upsilon^{(t)}_{i} = 1
\label{eq:sum_i_upsilon_constraint}
\end{equation}
and $\upsilon^{(t)}_{i} \ge 0$ for all $t \in I_c$, $i \in U_t^{-}$.

Solving the constrained optimization (\ref{eq:W_defn_for_Lagrange}) using Lagrange multipliers yields
\begin{equation}
\upsilon_i^{(t)} = \frac{A_i^{(t)} \gamma_t}{\delta} e^{-i\lambda} \quad \forall t \in I_c, i \in U_t^{-} \; ,
\label{eq:lagrange_mult_CN1}
\end{equation}
where $\lambda$ is the Lagrange multiplier. Substituting (\ref{eq:lagrange_mult_CN1}) into (\ref{eq:sum_i_upsilon_constraint}) and defining $z=e^{-\lambda}$ yields
\begin{equation}
\sum_{t \in I_c} \sum_{i \in U_t^-} i A_i^{(t)} \gamma_t z^i = \delta \, .
\label{eq:constraint_1}
\end{equation}
We may write this as
\begin{equation}
\sum_{t \in X_c} r A_r^{(t)} \gamma_t z^r + \sum_{t \in I_c} \sum_{i \in U_t^-\backslash\{r\}} i A_i^{(t)} \gamma_t z^i= \delta
\label{eq:constraint_2}
\end{equation}
from which we obtain (since all coefficients are positive and $z = e^{-\lambda} > 0$)
\[
\sum_{t \in X_c} r A_r^{(t)} \gamma_t z^r \leq \delta
\]
and therefore
\begin{equation}\label{eq:z_bound}
z \leq \frac{1}{\left(\sum_{t \in X_c} r A_r^{(t)} \gamma_t\right)^{1/r}} \, \delta^{1/r}
\end{equation}
is valid for all $\delta>0$. Thus
\begin{equation}
z \sim O(\delta^{1/r}) \; .
\label{eq:z_O_inverse_r}
\end{equation}
Recalling the definition $z = e^{-\lambda}$, (\ref{eq:lagrange_mult_CN1}) and (\ref{eq:z_O_inverse_r}) together imply that 
\begin{equation}
\upsilon^{(t)}_{i} \sim O\left(\delta^{i/r-1}\right) \quad \forall t \in I_c, i \in U_t^{-} \; .
\label{eq:O_notation}
\end{equation}

Next, since the value of $\bldupsilon$ which achieves the maximum in (\ref{eq:W_defn_for_Lagrange}) satisfies (\ref{eq:lagrange_mult_CN1}), we may develop (\ref{eq:W_defn_for_Lagrange}) as
\begin{eqnarray}
W & = & \delta \sum_{t \in I_c} \sum_{i \in U_t^-} \upsilon_i^{(t)} ( 1 + i\lambda ) \\
& = & \delta \left( \sum_{t \in I_c} \sum_{i \in U_t^-} \upsilon_i^{(t)} + \lambda \right) \label{eq:W_CN}
\end{eqnarray}
where in the second line we have used the constraint (\ref{eq:sum_i_upsilon_constraint}). 
 
Now, the constraint (\ref{eq:sum_i_upsilon_constraint}) may be written as
\[
1 = \sum_{t \in I_c} \sum_{i \in U_t^-} r \upsilon_i^{(t)} + \sum_{t,i \; :\; i > r} (i-r) \upsilon_i^{(t)} 
\]
so
\begin{equation}
\sum_{t \in I_c} \sum_{i \in U_t^-} \upsilon_i^{(t)} = \frac{1}{r} - \frac{1}{r} \sum_{t,i \; :\; i > r} (i-r) \upsilon_i^{(t)} \; .
\label{eq:sumupsilon}
\end{equation}
Also, we may write by (\ref{eq:lagrange_mult_CN1})
\[
\delta \upsilon_r^{(t)} e^{r \lambda} = A_r^{(t)} \gamma_t \; , \quad \forall t \in X_c \; .
\]
Multiplying by $r$ and summing over all $t \in X_c$ yields
\[
\delta e^{r \lambda} \sum_{t \in X_c} r \upsilon_r^{(t)}  = \sum_{t \in X_c} r A_r^{(t)} \gamma_t \; .
\]
Extracting $\lambda$ yields
\begin{eqnarray}
\lambda & = & \frac{1}{r} \log \left( \frac{ \sum_{t \in X_c} r A_r^{(t)} \gamma_t }{ \delta \sum_{t
\in X_c} r \upsilon_r^{(t)}} \right) \nonumber \\
& = & \frac{1}{r} \log \left( \frac{ \sum_{t \in X_c} r A_r^{(t)} \gamma_t }{ \delta } \right) -
\frac{1}{r} \log \left( \sum_{t \in X_c} r \upsilon_r^{(t)} \right) \nonumber \\
& = & \frac{1}{r} \log \left( \frac{ \sum_{t \in X_c} r A_r^{(t)} \gamma_t }{ \delta } \right) -
\frac{1}{r} \log \left( 1 - \sum_{t,i \; : \; i>r} i \upsilon_i^{(t)} \right) \; .
\label{eq:lambda_CN} 
\end{eqnarray}
where we have used (\ref{eq:sum_i_upsilon_constraint}) in the final line. Substituting
(\ref{eq:sumupsilon}) and (\ref{eq:lambda_CN}) back into (\ref{eq:W_CN}) yields
\begin{eqnarray*}
W & = & \left( \frac{\delta}{r} \right) \log \left( \frac{e \sum_{t \in X_c} r A_r^{(t)}
\gamma_t}{\delta} \right) - \left( \frac{\delta}{r} \right) \left[ \sum_{t,i \; :\; i > r} (i-r)
\upsilon_i^{(t)} + \log \left( 1 - \sum_{t,i \; : \; i>r} i \upsilon_i^{(t)} \right) \right] \\
& = & \left( \frac{\delta}{r} \right) \log \left( \frac{e C}{\delta \int \rho} \right) - \left(
\frac{\delta}{r} \right) \left[ \sum_{t,i \; :\; i > r} (i-r) \upsilon_i^{(t)} + \log \left( 1 -
\sum_{t,i \; : \; i>r} i \upsilon_i^{(t)} \right) \right] \\
& = & \left( \frac{\delta}{r} \right) \log \left( \frac{e C}{\delta \int \rho} \right) +
O(\delta^{\bar{r}/r})
\end{eqnarray*}
where in the second line we have used (\ref{eq:C_definition}), and in the final line we have used (\ref{eq:O_notation}) and the fact that $\log(1 + \alpha) \sim O(\alpha)$ (also recall the definition (\ref{eq:rbar_definition})). 

Substituting this expression for $W$ into (\ref{eq:N_c_max}) while recalling the $O(\delta^2)$ term in the expression for $W$, we have that as $m\rightarrow \infty$ 
\begin{equation}
N_c(\delta m) \asympequalm \exp \left\{ m \left[ \frac{\delta}{r} \log \left( \frac{e C}{\delta \int \rho} \right) + O \left(\delta^{\min\{\bar{r}/r,2\}}\right) \right] \right\} \; .
\label{eq:number_of_check_valid_assign} 
\end{equation}
Note that~(\ref{eq:number_of_check_valid_assign}) generalizes \cite[eqn. (30)]{Di_Richardson_Urbanke} to the case of a generalized \ac{CN} set.
 
\subsection{Number of variable-valid split assignments of split weight $(\tau n, \sigma n)$ over $\gamma n$ \acp{VN} of type $t \in I_v$}
Consider $\gamma n$ \acp{VN} of the same type $t \in I_v$. We now evaluate the number of variable-valid split assignments (over these \acp{VN}) of split weight $(\tau n, \sigma n)$. Using generating functions, this is given by\footnote{We use the following result on bivariate generating functions \cite{Wilf}. Let $a_{i,j}$ be the number of ways of obtaining an outcome $(i,j)\in\mathbb{Z}^2$ in experiment $\cA$, and let $b_{k,l}$ be the number of ways of obtaining an outcome $(k,l)\in\mathbb{Z}^2$ in experiment $\cB$. Also let $c_{p,q}$ be the number of ways of obtaining an outcome $((i,j),(k,l))$ in the combined experiment $(\cA, \cB)$ with sums $i+k=p$ and $j+l=q$. Then the generating functions $A(x,y)=\sum_{i,j} a_{i,j} x^i y^j$, $B(x,y)=\sum_{k,l} b_{k,l} x^k y^l$ and $C(x,y)=\sum_{p,q} c_{p,q} x^p y^q$ are related by $C(x,y) = A(x,y) B(x,y)$.}

\[
N_{v,t}^{(\gamma n)}(\tau n, \sigma n) = \coeff  \left[ \left( B^{(t)}(x,y) \right) ^{\gamma n}, x^{\tau n} y^{\sigma n} \right]
\]
where $\coeff [p(x,y), x^c y^d ]$ denotes the coefficient of $x^c y^d$ in the bivariate polynomial $p(x,y)$. We make use of the following result, the proof of which appears in \cite[Appendix A]{Burshtein_Miller}:
\medskip
\begin{lemma}
Let 
\[
B(x,y) = 1 + \sum_{u=1}^{k} \sum_{v=c}^{d} B_{u,v} x^u y^v 
\]
where $k \ge 1$ and $1 \le c \le d$, be a bivariate polynomial satisfying $B_{u,v} \ge 0$ for all $1 \le u \le k$, $c \le v \le d$. For fixed positive rational numbers $\xi$ and $\theta$, consider the set of positive integers $\ell$ such that $\xi \ell \in \mathbb{N}$, $\theta \ell \in \mathbb{N}$ and $\coeff [ (B(x,y) ) ^{\ell}, x^{\xi \ell}y^{\theta \ell} ] > 0$. Then either this set is empty, or has infinite cardinality; if $t$ is one such $\ell$, then so is $jt$ for every positive integer $j$. Assuming the latter case, the following limit is well defined and exists:
\begin{equation}
\lim_{\ell\rightarrow \infty} \frac{1}{\ell} \log \coeff \left[ \left( B(x,y) \right) ^{\ell}, x^{\xi \ell}y^{\theta \ell} \right] 
= \max_{\bldsmalleta} \sum_{(i,j) \in S} \eta_{i,j} \log \left( \frac{B_{i,j}}{\eta_{i,j}} \right)
\end{equation}
where $S = \{ (i,j) \in \mathbb{N}^2 \; : \; B_{i,j} > 0 \}$, $\bldeta = ( \eta_{i,j} )_{(i,j) \in S}$, and the maximization is subject to the constraints 
$\sum_{(i,j) \in S} \eta_{i,j} = 1$, $\sum_{(i,j) \in S} i \eta_{i,j} = \xi$, $\sum_{(i,j) \in S} j \eta_{i,j} = \theta$ and $\eta_{i,j} \ge 0$ for all $(i,j) \in S$.
\label{lemma:optimization_2D}
\end{lemma}
\medskip

Applying this lemma by substituting $B(x,y) = B^{(t)}(x,y)$, $\ell = \gamma n$, $\xi = \tau/\gamma$ and $\theta = \sigma/\gamma$, we obtain that with $\gamma$ fixed, as $n \rightarrow \infty$ 
\begin{eqnarray}
N_{v,t}^{(\gamma n)}(\tau n, \sigma n) = \coeff \left[ \left( B^{(t)}(x,y) \right) ^{\gamma n}, x^{\tau n} y^{\sigma n} \right] 
\label{eq:Nvt_tau_sigma_start} \\
\asympequaln \exp \left\{ n \gamma \max_{\bldsmalleta^{(t)}} \sum_{(i,j) \in S_t} \eta^{(t)}_{i,j} \log \left( \frac{B^{(t)}_{i,j}}{\eta^{(t)}_{i,j}} \right) \right\} 
\label{eq:Nvt_tau_sigma_mid} \\
\triangleq \exp \left\{ n X^{(\gamma)}_t(\tau, \sigma) \right\}
\label{eq:Nvt_tau_sigma_end}
\end{eqnarray}
where the maximization over $\bldeta^{(t)} = ( \eta^{(t)}_{i,j} )_{(i,j) \in S_t}$ is subject to the constraints $\sum_{(i,j) \in S_t} \eta^{(t)}_{i,j} = 1$, $\sum_{(i,j) \in S_t^{-}} i \eta^{(t)}_{i,j} = \tau / \gamma$, $\sum_{(i,j) \in S_t^{-}} j \eta^{(t)}_{i,j} = \sigma / \gamma$ and $\eta^{(t)}_{i,j} \ge 0$ for all $(i,j) \in S_t$ (recall that the sets $S_t$ and $S_t^{-}$ are given by (\ref{eq:St}) and (\ref{eq:St-})).

\subsection{Growth rate of the weight distribution of the irregular D-GLDPC code ensemble sequence}

Recall that the number of check-valid assignments of weight $\delta m$ is $N_c(\delta m)$; also, the total number of assignments of weight $\delta m$ is $\binom{E}{\delta m}$. Therefore, the probability that a randomly chosen assignment of weight $\delta m$ is check-valid is given by 
\[
P_{\mbox{\scriptsize valid}}(\delta m) = N_c(\delta m) \Big/ \binom{E}{\delta m} \; .
\]
Here we adopt the notation $\delta m = \beta n$; also we have $E = m / \int \rho = n / \int \lambda$. The binomial coefficient may be asymptotically approximated using the fact, based on Stirling's approximation, that as $n \rightarrow \infty$ \cite{Di_Richardson_Urbanke}
\[
\binom{\tau n}{\sigma n} \asympequaln \exp \left\{ n \left[ \sigma \log \left( \frac{e \tau}{\sigma} \right) + 
O(\sigma^2) \right] \right\}
\]
(valid for $0 < \sigma < \tau < 1$) which yields, in this case,
\[
\binom{n / \int \lambda}{\beta n} \asympequaln \exp \left\{ n \left[ \beta \log \left( \frac{e}{\beta \int \lambda} \right) + 
O(\beta^2) \right] \right\}
\]
as $n \rightarrow \infty$. Applying this together with the asymptotic expression (\ref{eq:number_of_check_valid_assign}), we find that as $n \rightarrow \infty$ (exploiting the fact that $\delta \int \rho = \beta \int \lambda$) 
\begin{equation}
P_{\mbox{\scriptsize valid}}(\beta n) \asympequaln \exp \{ n Y(\beta)\}
\label{eq:Pvalid_limit}
\end{equation}
where 
\begin{equation}
Y(\beta) = \frac{\beta}{r} \log \left( \frac{e C}{\beta \int \lambda} \right) - \beta \log \left( \frac{e}{\beta \int \lambda} \right) + O \left(\beta^{\min\{\bar{r}/r,2\}}\right) \; .
\label{eq:Y_of_beta}
\end{equation}

Next, we note that the expected number of D-GLDPC codewords of weight $\alpha n$ in the ensemble $\cM_n$ is equal to the sum over $\beta$ of the expected number of split assignments of split weight $(\alpha n, \beta n)$ which are both check-valid and variable-valid, denoted $N^{v,c}_{\alpha n, \beta n}$:
\[
\mathbb{E}_{\cM_n} \left[ N_{\alpha n} \right] = \sum_{\beta} \mathbb{E}_{\cM_n} [ N^{v,c}_{\alpha n, \beta n} ] \; .
\]
This may then be expressed as  
\begin{equation*}
\mathbb{E}_{\cM_n} \left[ N_{\alpha n} \right] =
\sum_{\beta} P_{\mbox{\scriptsize valid}}(\beta n) \sum_{\substack{\sum \alpha_t = \alpha \\ \sum \beta_t = \beta}} \left[ \prod_{t \in I_v} N_{v,t}^{(\delta_t n)}(\alpha_t n, \beta_t n) \right]
\end{equation*}
where the fraction $\delta_t$ of \acp{VN} of type $t \in I_v$ is given by (\ref{eq:delta_t_definition})
and the second sum is over all partitions of $\alpha$ and $\beta$ into $n_v$ elements, i.e., we have $\alpha_t, \beta_t \ge 0$ for all $t \in I_v$, and $\sum_{t \in I_v} \alpha_t = \alpha$, $\sum_{t \in I_v} \beta_t = \beta$. 

Now, using (\ref{eq:Nvt_tau_sigma_start})-(\ref{eq:Nvt_tau_sigma_end}), as $n \rightarrow \infty$ we have for each $t \in I_v$
\begin{equation*}
N_{v,t}^{(\delta_t n)}(\alpha_t n, \beta_t n) \asympequaln \exp \left\{ n X^{(\delta_t)}_t(\alpha_t, \beta_t) \right\} \; ,
\end{equation*}
where, for each $t \in I_v$,
\begin{equation}
X^{(\delta_t)}_t(\alpha_t, \beta_t) = \delta_t \max_{\bldsmalleta^{(t)}} \sum_{(i,j) \in S_t} \eta^{(t)}_{i,j} \log \left( \frac{B^{(t)}_{i,j}}{\eta^{(t)}_{i,j}} \right)
\label{eq:X_function}
\end{equation}
and the maximization over $\bldeta^{(t)} = ( \eta^{(t)}_{i,j} )_{(i,j) \in S_t}$ is subject to the constraints 
\begin{equation}
\sum_{(i,j) \in S_t} \eta^{(t)}_{i,j} = 1
\label{eq:eta_sum_constraint}
\end{equation}
\begin{equation}
\sum_{(i,j) \in S_t^{-}} i \eta^{(t)}_{i,j} = \alpha_t / \delta_t
\label{eq:sum_xi_constraint}
\end{equation}
\begin{equation}
\sum_{(i,j) \in S_t^{-}} j \eta^{(t)}_{i,j} = \beta_t / \delta_t
\label{eq:sum_theta_constraint}
\end{equation}
and
\begin{equation}
\eta^{(t)}_{i,j} \ge 0 \quad \forall (i,j) \in S_t \; .
\label{eq:nonnegative_eta_constraint}
\end{equation}
Therefore, recalling (\ref{eq:Pvalid_limit}), we have that as $n \rightarrow \infty$, 
\begin{equation}
\mathbb{E}_{\cM_n} \left[ N_{\alpha n} \right] \asympequaln
\sum_{\beta} \sum_{\substack{\sum \alpha_t = \alpha \\ \sum \beta_t = \beta}}
\exp \left\{ n \left[ \sum_{t \in I_v} X^{(\delta_t)}_t(\alpha_t, \beta_t) + Y(\beta) \right] \right\} \; .
\label{eq:sum_of_exp}
\end{equation}
Next, for each $t \in I_v$ we define
\[
F_t(\bldeta^{(t)}) = \eta^{(t)}_{0,0} \log \left( \frac{1}{\eta^{(t)}_{0,0}} \right) - \sum_{(i,j) \in S_t^{-}} \eta^{(t)}_{i,j} \; .
\]
Note that the expression (\ref{eq:sum_of_exp}) is dominated as $n \rightarrow \infty$ by the term which maximizes the argument of the exponential. Thus using~(\ref{eq:Y_of_beta}) and~(\ref{eq:X_function}) we may write
\begin{multline}
G(\alpha) = \max_{\beta} \max_{\substack{\sum \alpha_t = \alpha \\ \sum \beta_t = \beta}} \Bigg\{ \sum_{t \in I_v} \delta_t \max_{\bldsmalleta^{(t)}} \Bigg[ \sum_{(i,j) \in S_t^{-}} \eta^{(t)}_{i,j} \log \left( \frac{e B^{(t)}_{i,j}}{\eta^{(t)}_{i,j}} \right) + F_t(\bldeta^{(t)}) \Bigg] \\
+ \frac{\beta}{r} \log \left( \frac{e C}{\beta \int \lambda} \right) - \beta \log \left( \frac{e}{\beta \int \lambda} \right) + O \left(\beta^{\min\{\bar{r}/r,2\}}\right) \Bigg\}
\label{eq:growth_rate_with_O_notation}
\end{multline}
where the maximization over $\bldeta^{(t)} = ( \eta^{(t)}_{i,j} )_{(i,j) \in S_t^{-}}$ (for each $t \in I_v$) is subject to constraints (\ref{eq:sum_xi_constraint}) and (\ref{eq:sum_theta_constraint}) together with $\eta^{(t)}_{i,j} \ge 0$ for all $(i,j) \in S_t^{-}$.

We next have the following lemma.
\medskip
\begin{lemma}
The expression $\sum_{t \in I_v} \delta_t F_t(\bldeta^{(t)})$ is $O(\alpha^2)$ for any $\bldeta^{(t)}$ satisfying the optimization constraints~(\ref{eq:eta_sum_constraint})-(\ref{eq:nonnegative_eta_constraint}).
\label{lemma:O2_term}
\end{lemma}
\medskip
The proof of this lemma follows the same lines as the proof of Lemma \ref{lemma:O2_term_check}, and is therefore omitted. It follows from Lemma \ref{lemma:O2_term} that the expression $\sum_{t \in I_v} \delta_t F_t(\bldeta^{(t)})$ is $O(\alpha^2)$ for the maximizing $\bldeta^{(t)}$. Also, since $\beta/\alpha$ is bounded between two positive constants, any expression which is $O(\beta^{\kappa})$ must necessarily also be $O(\alpha^{\kappa})$ (where $\kappa>0$). Therefore
\begin{multline*}
G(\alpha) = \max_{\beta} \max_{\substack{\sum \alpha_t = \alpha \\ \sum \beta_t = \beta}} \Bigg[ \sum_{t \in I_v} \delta_t \max_{\bldsmalleta^{(t)}} \sum_{(i,j) \in S_t^{-}} \eta^{(t)}_{i,j} \log \left( \frac{e B^{(t)}_{i,j}}{\eta^{(t)}_{i,j}} \right) \\
+ \frac{\beta}{r} \log \left( \frac{e C}{\beta \int \lambda} \right) - \beta \log \left( \frac{e}{\beta \int \lambda} \right) \Bigg] + O \left(\alpha^{\min\{\bar{r}/r,2\}}\right)
\end{multline*}
where the optimization is (as before) subject to the constraints (\ref{eq:sum_xi_constraint}) and (\ref{eq:sum_theta_constraint}) together with $\eta^{(t)}_{i,j} \ge 0$ for all $(i,j) \in S_t^{-}$. In what follows, for convenience of presentation we shall temporarily omit the $O \left(\alpha^{\min\{\bar{r}/r,2\}}\right)$ term in the expression for the growth rate.

Next we make the substitution $\gamma^{(t)}_{i,j} = \delta_t \eta^{(t)}_{i,j}$ for all $t \in I_v$, $(i,j) \in S_t^{-}$. This yields
\begin{equation*}
G(\alpha) = \max_{\beta} \max_{\substack{\sum \alpha_t = \alpha \\ \sum \beta_t = \beta}} \Bigg[ \sum_{t \in I_v} \max_{\bldsmallgamma^{(t)}} \sum_{(i,j) \in S_t^{-}} \gamma^{(t)}_{i,j} \log \left( \frac{e B^{(t)}_{i,j} \delta_t}{\gamma^{(t)}_{i,j}} \right)
+ \frac{\beta}{r} \log \left( \frac{e C}{\beta \int \lambda} \right) - \beta \log \left( \frac{e}{\beta \int \lambda} \right) \Bigg]
\end{equation*}
where the maximization over $\bldgamma^{(t)} = ( \gamma^{(t)}_{i,j} )_{(i,j) \in S_t^{-}}$ (for each $t \in I_v$) is subject to the constraints $\sum_{(i,j) \in S_t^{-}} i \gamma^{(t)}_{i,j} = \alpha_t$, $\sum_{(i,j) \in S_t^{-}} j \gamma^{(t)}_{i,j} = \beta_t$, and $\gamma^{(t)}_{i,j} \ge 0$ for all $(i,j) \in S_t^{-}$.
We observe that this maximization may be recast as 
\begin{equation*}
G(\alpha) = \max_{\bldsmallgamma} \Bigg[ \sum_{t \in I_v} \sum_{(i,j) \in S_t^{-}} \gamma^{(t)}_{i,j} \log \left( \frac{e B^{(t)}_{i,j} \delta_t}{\gamma^{(t)}_{i,j}} \right) \\
+ \frac{\beta(\bldgamma)}{r} \log \left( \frac{e C}{\beta(\bldgamma) \int \lambda} \right) - \beta(\bldgamma) \log \left( \frac{e}{\beta(\bldgamma) \int \lambda} \right) \Bigg]
\end{equation*}
where the maximization, which is now over $\bldgamma = (\gamma^{(t)}_{i,j})_{t \in I_v, (i,j) \in S_t^{-}}$, is subject to the constraints 
\[
\sum_{t \in I_v} \sum_{(i,j) \in S_t^{-}} i \gamma^{(t)}_{i,j} = \alpha
\]
and $\gamma^{(t)}_{i,j} \ge 0$ for all $t \in I_v$, $(i,j) \in S_t^{-}$, and where
\[
\beta(\bldgamma) = \sum_{t \in I_v} \sum_{(i,j) \in S_t^{-}} j \gamma^{(t)}_{i,j} \; .
\]
Making the substitution $\nu^{(t)}_{i,j} = \gamma^{(t)}_{i,j} / \alpha$ for all $t \in I_v$, $(i,j) \in S_t^{-}$, we obtain
\begin{multline}
G(\alpha) = \alpha \max_{\bldsmallnu} \Bigg[ \sum_{t \in I_v} \sum_{(i,j) \in S_t^{-}} \nu^{(t)}_{i,j} \log \left( \frac{e B^{(t)}_{i,j} \delta_t}{\alpha \nu^{(t)}_{i,j}} \right) \\
+ \frac{z(\bldnu)}{r} \log \left( \frac{e C}{\alpha z(\bldnu) \int \lambda} \right) - z(\bldnu) \log \left( \frac{e}{\alpha z(\bldnu) \int \lambda} \right) \Bigg]
\label{eq:G_defn}
\end{multline}
where the maximization over $\bldnu = (\nu^{(t)}_{i,j})_{t \in I_v, (i,j) \in S_t^{-}}$ is subject to the constraint 
\begin{equation}
\sum_{t \in I_v} \sum_{(i,j) \in S_t^{-}} i \nu^{(t)}_{i,j} = 1 \; ,
\label{eq:LM_constraint}
\end{equation}
as well as $\nu^{(t)}_{i,j} \ge 0$ for all $t \in I_v$, $(i,j) \in S_t^{-}$, and where 
\begin{equation}
z(\bldnu) \triangleq \sum_{t \in I_v} \sum_{(i,j) \in S_t^{-}} j \nu^{(t)}_{i,j} \; .
\label{eq:z_defn}
\end{equation}
By solving (\ref{eq:G_defn}) directly using Lagrange multipliers, one shows that 
\begin{equation}
\log \left( \frac{B^{(t)}_{i,j} \delta_t}{\alpha \nu^{(t)}_{i,j}} \right) + \frac{j}{r} \log \left( \frac{C}{\alpha z(\bldnu) \int \lambda} \right) - j \log \left( \frac{1}{\alpha z(\bldnu) \int \lambda} \right) = \lambda i
\label{eq:Lagrange_soln_1}
\end{equation}
holds for all $t \in I_v$, $(i,j) \in S_t^{-}$; here $\lambda$ denotes the Lagrange multiplier. Substituting (\ref{eq:Lagrange_soln_1}) back into (\ref{eq:G_defn}) yields 
\begin{eqnarray}
G(\alpha) & = & \alpha \Bigg( \sum_{t \in I_v} \sum_{(i,j) \in S_t^{-}} \nu^{(t)}_{i,j} \left[ 1 + \lambda i - \frac{j}{r} \log \left( \frac{C}{\alpha z(\bldnu) \int \lambda} \right) + j \log \left( \frac{1}{\alpha z(\bldnu) \int \lambda} \right) \right] \nonumber \\
& + & \frac{z(\bldnu)}{r} \log \left( \frac{e C}{\alpha z(\bldnu) \int \lambda} \right) - z(\bldnu) \log \left( \frac{e}{\alpha z(\bldnu) \int \lambda} \right) \Bigg) \nonumber \\
& = & \alpha \left[ \sum_{t \in I_v} \sum_{(i,j) \in S_t^{-}} \nu^{(t)}_{i,j} + \lambda + \frac{z(\bldnu)}{r} - z(\bldnu) \right] \nonumber \\
& = & \alpha \left[ \lambda + \sum_{t \in I_v} \sum_{(i,j) \in S_t^{-}} \nu^{(t)}_{i,j} \left( 1 + \frac{j}{r} - j \right) \right] \nonumber \\
& = & \alpha \left[ \lambda - \frac{1}{\psi} \sum_{t \in I_v} \sum_{(i,j) \in S_t^{-}} \nu^{(t)}_{i,j} \left( j-\psi \right) \right] \; ,
\label{eq:G1}
\end{eqnarray}
where in the second line we have used (\ref{eq:LM_constraint}) and (\ref{eq:z_defn}). Note that (\ref{eq:Lagrange_soln_1}) may be rearranged as
\begin{equation}\label{eq:pre_omega_bound}
\alpha \nu^{(t)}_{i,j} = B^{(t)}_{i,j} \delta_t C^{j/r} \left( \alpha z(\bldnu) \int \lambda \right)^{j/\psi} \omega^i 
\end{equation}
where $\omega = e^{-\lambda}$ is a positive real number.
Substituting this solution into (\ref{eq:z_defn}) (i.e. into the definition of $z(\bldnu)$) and using \eqref{eq:delta_t_definition} yields
\begin{equation}
\sum_{t \in I_v} \frac{\lambda_t}{q_t} \sum_{(i,j) \in S_t^{-}} j B^{(t)}_{i,j} C^{j/r} \left( \alpha z(\bldnu) \int \lambda \right)^{\frac{j-\psi}{\psi}} \omega^i = 1 \; , 
\label{eq:omega_eqn_1}
\end{equation}
and similarly the constraint (\ref{eq:LM_constraint}) may be written as
\begin{equation}
\sum_{t \in I_v} \frac{\lambda_t}{q_t} \sum_{(i,j) \in S_t^{-}} i B^{(t)}_{i,j} C^{j/r} \left( \alpha z(\bldnu) \int \lambda \right)^{\frac{j-\psi}{\psi}} \omega^i = \frac{1}{z(\bldnu)} \; .
\label{eq:omega_eqn_2}
\end{equation}

We next proceed by proving an upper bound on all terms $\nu^{(t)}_{i,j}$ such that $(t,i,j)$ does not lie in the dominant set discussed in Section \ref{subsection:graphical_interpretation}, i.e., on all terms $\nu^{(t)}_{i,j}$ where $t \in I_v$, $(i,j) \in S_t^{-}$ and $T_{i,j} > T$. To this end, fix $s \in I_v$, $(k,l) \in S_s^{-}$ with $T_{k,l} = T$, and consider arbitrary $t \in I_v$, $(i,j) \in S_t^{-}$ (for which we have $T_{i,j} \ge T$). Then applying (\ref{eq:Lagrange_soln_1}) in the two pertinent cases yields
\[
\lambda = \frac{1}{i} \log \left[ \frac{B^{(t)}_{i,j} \delta_t C^{j/r}}{\alpha \nu^{(t)}_{i,j}} \cdot \left( \alpha z(\bldnu) \int \lambda \right)^{j/\psi} \right]
\]
and
\[
\lambda = \frac{1}{k} \log \left[ \frac{B^{(s)}_{k,l} \delta_s C^{l/r}}{\alpha \nu^{(s)}_{k,l}} \cdot \left( \alpha z(\bldnu) \int \lambda \right)^{l/\psi} \right] \; .
\]
Equating these two expressions for $\lambda$ we obtain
\[
\frac{1}{i} \log \nu^{(t)}_{i,j} - \frac{1}{k} \log \nu^{(s)}_{k,l} = \left(
\frac{T_{i,j} - T}{\psi} \right) \log \alpha +
F_{i,j}^{(t)}(\alpha) - F_{k,l}^{(s)}(\alpha)
\]
where we define the function
\begin{equation}
F_{i,j}^{(t)}(\alpha) = \frac{1}{i} \log \left[ B^{(t)}_{i,j} \delta_t C^{j/r}
\left( z(\bldnu) \int \lambda \right)^{j/\psi} \right] \; .
\label{eq:Fijt_definition}
\end{equation}
for every $t \in I_v$, $(i,j) \in S_t^{-}$ (it is easy to check that this is indeed a function of $\alpha$). Note that $z(\bldnu)$ is bounded above and below as $\alpha \rightarrow 0$; this may be easily shown since \eqref{eq:LM_constraint} and \eqref{eq:z_defn} imply
\[
0 < \left( \min_{(i,j) \in S^{-}} i\right) \sum_{t \in I_v} \sum_{(i,j) \in S_t^{-}} \nu^{(t)}_{i,j} \le 1 \le \left( \max_{(i,j) \in S^{-}} i\right) \sum_{t \in I_v} \sum_{(i,j) \in S_t^{-}} \nu^{(t)}_{i,j} 
\] 
and
\[
0 < \left( \min_{(i,j) \in S^{-}} j\right) \sum_{t \in I_v} \sum_{(i,j) \in S_t^{-}} \nu^{(t)}_{i,j} \le z(\bldnu) \le \left( \max_{(i,j) \in S^{-}} j\right) \sum_{t \in I_v} \sum_{(i,j) \in S_t^{-}} \nu^{(t)}_{i,j} 
\]
respectively. Thus it follows that for any $t \in I_v$, $(i,j) \in S_t^{-}$, the function $F_{i,j}^{(t)}(\alpha)$ given by \eqref{eq:Fijt_definition} is bounded above and below as $\alpha \rightarrow 0$. Therefore
\begin{equation}\label{eq:nu_ratio}\frac{(\nu^{(t)}_{i,j})^{1/i}}{( \nu^{(s)}_{k,l} )^{1/k}} = \frac{\Gamma_{i,j}^{(t)}(\alpha)}{\Gamma_{k,l}^{(s)}(\alpha)}\,
\alpha^{\frac{T_{i,j} - T}{\psi}}\end{equation}
\noindent where
$\Gamma_{i,j}^{(t)}(\alpha) = \exp(F_{i,j}^{(t)}(\alpha))>0$ for every $t \in I_v$, $(i,j) \in S_t^{-}$. 
We next write \eqref{eq:nu_ratio} as
\begin{equation} 
\nu^{(t)}_{i,j} = (\nu^{(s)}_{k,l})^{i/k} \Delta_{i,j,k,l}^{(t,s)}(\alpha)\,\alpha^{\frac{T_{i,j} - T}{\psi}i}
\label{eq:nu_explicit}
\end{equation} 
where for conciseness we have defined
\[
\Delta_{i,j,k,l}^{(t,s)}(\alpha) = \left(\frac{\Gamma_{i,j}^{(t)}(\alpha)}{\Gamma_{k,l}^{(s)}
(\alpha)}\right)^i > 0 \; .
\]
Substituting into \eqref{eq:LM_constraint} we obtain
\[
\sum_{t \in I_v}\sum_{(i,j)\in S_t^-} i (\nu^{(s)}_{k,l})^{i/k} 
\Delta_{i,j,k,l}^{(t,s)}(\alpha)\,\alpha^{\frac{T_{i,j} - T}{\psi}i} = 1 \, ,
\]
which in particular implies (since all terms involved are nonnegative, and $T_{i,j}=T$ for all $t \in Y_v$, $(i,j) \in P_t$)
\[
\sum_{t \in Y_v}\sum_{(i,j)\in P_t} i (\nu^{(s)}_{k,l})^{i/k}
\Delta_{i,j,k,l}^{(t,s)}(\alpha) \leq 1\, .
\]
If we now define $\hat{i}=\max\{i : (i,j) \in P_t \textrm{ for some } t \in Y_v\}$, the
previous inequality leads to
\[
\nu^{(s)}_{k,l} \leq \left( \frac{1}{\sum_{t \in Y_v}\sum_{(i,j)\in P_t} i
\Delta_{i,j,k,l}^{(t,s)}(\alpha)} \right) ^{k/\hat{i}}\, .
\]
Since, for every triple $(t,i,j)$, $F_{i,j}^{(t)}(\alpha)$ is bounded above and below, so are
$\Gamma_{i,j}^{(t)}(\alpha)$ and $\Delta_{i,j,k,l}^{(t,s)}(\alpha)$. Therefore, the previous
inequality implies
\begin{equation}
\nu^{(s)}_{k,l} \sim O(1) \, ,
\label{eq:nuijt_big_O}
\end{equation}
which holds for any $s \in Y_v$, $(k,l) \in P_s$, i.e., for any triple $(s,k,l)$ lying in the dominant set. Recalling \eqref{eq:nu_explicit}, for general $\nu_{i,j}^{(t)}$ we have
\begin{equation}
\nu_{i,j}^{(t)} \sim O(\alpha^{\frac{T_{i,j}-T}{\psi}i}) \qquad \textrm{for all }t\in I_v, \,(i,j) \in S_t^- \, .
\label{eq:nukls_big_O}
\end{equation}

Next observe that \eqref{eq:nukls_big_O} may be used to upper bound $\omega=e^{-\lambda}$. In fact, it may be seen that applying \eqref{eq:pre_omega_bound} for any triple $(t,i,j)$ and taking into account \eqref{eq:nukls_big_O} leads to
\[
\omega^i = \frac{1}{B_{i,j}^{(t)}\delta_t C^{j/r}}\cdot\frac{\alpha\nu_{i,j}^{(t)}}{(\alpha z(\bldnu) \int \lambda)^{j/\psi}} \sim O(\alpha^{-i\frac{T}{\psi}})
\]
i.e.,
\begin{equation}
\omega \sim O(\alpha^{-\frac{T}{\psi}}) \, .
\label{eq:omega_bound}
\end{equation}

Next, \eqref{eq:nukls_big_O} implies that 
\begin{eqnarray}
\frac{1}{\psi} \sum_{t \in I_v} \sum_{(i,j) \in S_t^{-}} (j-\psi) \nu^{(t)}_{i,j} & = & \frac{1}{\psi} \left[ \sum_{t \in I_v} \sum_{(i,j) \in S_t^{-}} iT \nu^{(t)}_{i,j} + \sum_{t \in I_v} \sum_{(i,j) \in S_t^{-}} i(T_{i,j}-T) \nu^{(t)}_{i,j} \right] \nonumber \\
& = & \frac{T}{\psi} + \frac{1}{\psi} \sum_{t \in I_v} \sum_{(i,j):T_{i,j}>T} i(T_{i,j}-T) \nu^{(t)}_{i,j} \nonumber \\
& = & \frac{T}{\psi} + O \left(\alpha^\frac{\chi}{\psi}\right) \; .
\label{eq:Part_I_solution}
\end{eqnarray}
where in the second line we have used \eqref{eq:LM_constraint}, and in the final line we have used \eqref{eq:nukls_big_O} and also recalled the definition \eqref{eq:chi_definition} of the parameter $\chi$.

Next note that, recalling \eqref{eq:P1x_definition} and \eqref{eq:P2x_definition},
equations (\ref{eq:omega_eqn_1})
and (\ref{eq:omega_eqn_2}) may be written as
\begin{equation}
Q_1\left(y\right) + \mathsf{f}_1(\alpha) = 1
\label{eq:Q1_final_system}
\end{equation}
and
\begin{equation}
Q_2\left(y\right) + \mathsf{f}_2(\alpha) = \frac{1}{z(\bldnu)}
\label{eq:Q2_final_system}
\end{equation}
respectively, where we define
\begin{equation}
y = \left( e\alpha z(\bldnu) \right)^{T/\psi}\omega \; ,
\label{eq:y_definition}
\end{equation}
\[
\mathsf{f}_1(\alpha) = \sum_{t \in I_v} \frac{\lambda_t}{q_t} \sum_{(i,j):T_{i,j}>T} jB_{i,j}^{(t)}C^{j/r} \left[ \left(\alpha z(\bldnu)\int\!\lambda\right)^{T_{i,j}/\psi}\omega \right]^i
\]
and
\[
\mathsf{f}_2(\alpha) = \sum_{t \in I_v} \frac{\lambda_t}{q_t} \sum_{(i,j):T_{i,j}>T} iB_{i,j}^{(t)}C^{j/r} \left[ \left(\alpha z(\bldnu)\int\!\lambda\right)^{T_{i,j}/\psi}\omega \right]^i \, .
\]
Note that since $z(\bldnu)$ is bounded below as $\alpha \rightarrow 0$, from \eqref{eq:omega_bound} we have $\mathsf{f}_1(\alpha) \sim O(\alpha^{\frac{\chi}{\psi}})$ and $\mathsf{f}_2(\alpha) \sim O(\alpha^{\frac{\chi}{\psi}})$. Recalling that $\omega=e^{-\lambda}$, from \eqref{eq:y_definition} we obtain the following
expression for the Lagrange multiplier $\lambda$:
\begin{equation}
\lambda = \frac{T}{\psi}+\frac{T}{\psi}\log\alpha+\frac{T}{\psi}\log z(\bldnu)-\log y\, .
\label{eq:lambda_in_terms_of_y}
\end{equation}
From \eqref{eq:Q1_final_system} and \eqref{eq:Q2_final_system} this latter expression may be
written as
\[
\lambda=\frac{T}{\psi}+\frac{T}{\psi}\log\alpha-\frac{T}{\psi}\log\left[Q_2(Q_1^{-1}
(1-\mathsf{f_1(\alpha)})) + \mathsf{f}_2(\alpha) \right] -\log \left[Q_1^{-1}
(1-\mathsf{f_1(\alpha)})\right]\, .
\]
Using the Taylor series of $Q_1^{-1}(1+x)$ around $x=0$, we have
\begin{equation}
Q_1^{-1}(1-\mathsf{f_1(\alpha)})=Q_1^{-1}(1) + \mathsf{g}_1(\alpha) \; ,
\label{eq:Q1_inverse_1_minus_f1}
\end{equation}
where
\[
\mathsf{g}_1(\alpha)=\frac{{\rm d}Q_1^{-1}}{{\rm d}x}(1)\cdot\mathsf{f}_1(\alpha) + O(\mathsf{f}_1^2(\alpha))\, .
\]
Note that since $\mathsf{f}_1(\alpha) \sim O(\alpha^\frac{\chi}{\psi})$, we also have $\mathsf{g}_1(\alpha) \sim O(\alpha^\frac{\chi}{\psi})$. Substituting the obtained expression \eqref{eq:Q1_inverse_1_minus_f1} for $Q_1^{-1}(1-\mathsf{f_1(\alpha)})$ into the previous expression \eqref{eq:lambda_in_terms_of_y} for $\lambda$ we obtain
\[
\lambda=\frac{T}{\psi}+\frac{T}{\psi}\log\alpha-\frac{T}{\psi}\log\left[Q_2(Q_1^{-1}(1) +
\mathsf{g}_1(\alpha))) + \mathsf{f}_2(\alpha) \right] -\log \left[Q_1^{-1}(1) +
\mathsf{g}_1(\alpha)\right]\, .
\]
We may now develop $Q_2(Q_1^{-1}(1) + \mathsf{g}_1(\alpha)))$ using the Taylor series for
$Q_2(Q_1^{-1}(1) + x)$ around $x=0$ and $\log \left[Q_1^{-1}(1) +
\mathsf{g}_1(\alpha)\right]$ using the Taylor series for $\log(Q_1^{-1}(1) + x)$ around
$x=0$. We obtain
\[
Q_2(Q_1^{-1}(1) + \mathsf{g}_1(\alpha))) = Q_2(Q_1^{-1}(1)) + \mathsf{g}_2(\alpha)
\]
where
\[
\mathsf{g}_2(\alpha)=\frac{{\rm d}Q_2}{{\rm d}x} (Q_1^{-1}(1)) \cdot
\mathsf{g}_1(\alpha)+O(\mathsf{g}_1^2(\alpha))
\]
and
\[
\log\left[Q_1^{-1}(1)+\mathsf{g}_1(\alpha)\right]=\log Q_1^{-1}(1) + \mathsf{h}_1(\alpha)
\]
where $$\mathsf{h}_1(\alpha)=\frac{1}{Q_1^{-1}(1)}\mathsf{g}_1(\alpha)+O(\mathsf{g}_1^2(\alpha))\,.$$
Again, note that since $\mathsf{g}_1(\alpha)\sim O(\alpha^\frac{\chi}{\psi})$, we have
$\mathsf{g}_2(\alpha) \sim O(\alpha^\frac{\chi}{\psi})$ and $\mathsf{h}_1(\alpha) \sim
O(\alpha^{\frac{\chi}{\psi}})$. Therefore, we have
\begin{equation*}
\lambda=\frac{T}{\psi}+\frac{T}{\psi}\log\alpha-\frac{T}{\psi}\log\left[Q_2(Q_1^{-1}(1)) +
\mathsf{g}_2(\alpha) + \mathsf{f}_2(\alpha) \right] +\log \frac{1}{Q_1^{-1}(1)} -
\mathsf{h}_1(\alpha)\, .
\end{equation*}
Finally, we develop $\log\left[Q_2(Q_1^{-1}(1)) + \mathsf{g}_2(\alpha) + \mathsf{f}_2(\alpha)
\right]$ using the Taylor series for $\log\left[Q_2(Q_1^{-1}(1)) + x \right]$ around
$x=0$. We obtain
\[
\log\left[Q_2(Q_1^{-1}(1)) + \mathsf{g}_2(\alpha) + \mathsf{f}_2(\alpha) \right] = \log
Q_2(Q_1^{-1}(1)) + \mathsf{h}_2(\alpha)
\]
where
\[
\mathsf{h}_2(\alpha)=\frac{1}{Q_2(Q_1^{-1}(1))} (\mathsf{g}_2(\alpha)+\mathsf{f}_2(\alpha)) +
O((\mathsf{g}_2(\alpha)+\mathsf{f}_2(\alpha))^2)\, .
\]
Using this expression we obtain
\[
\lambda=\frac{T}{\psi}+\frac{T}{\psi}\log\alpha + \frac{T}{\psi}\log\frac{1}{Q_2(Q_1^{-1}(1))} + \log \frac{1}{Q_1^{-1}(1)} - \frac{T}{\psi} \mathsf{h}_2(\alpha) -
\mathsf{h}_1(\alpha)\, .
\]
Since $\mathsf{g}_2(\alpha) \sim O(\alpha^\frac{\chi}{\psi})$ and $\mathsf{f}_2(\alpha) \sim O(\alpha^\frac{\chi}{\psi})$, we have $\mathsf{h}_2(\alpha) \sim O(\alpha^\frac{\chi}{\psi})$; hence, we obtain
\begin{equation}
\lambda=\frac{T}{\psi}+\frac{T}{\psi}\log\alpha + \frac{T}{\psi}\log\frac{1}{Q_2(Q_1^{-1}(1))} + \log \frac{1}{Q_1^{-1}(1)} + O(\alpha^\frac{\chi}{\psi})\, .
\label{eq:Part_II_solution}
\end{equation}

Finally, substituting \eqref{eq:Part_I_solution} and \eqref{eq:Part_II_solution} into \eqref{eq:G1} and recalling the $O \left(\alpha^{\min\{\bar{r}/r,2\}}\right)$ term completes the proof of the theorem. 

\section{Conclusion}
\label{section:conclusion}
A compact expression for the asymptotic growth rate of the weight distribution of irregular D-GLDPC codes for small linear-weight codewords has been derived. Ensembles with check or variable node minimum distance greater than $2$ are shown to have good growth rate behavior, while for other ensembles an important parameter is identified which discriminates between good and bad growth rate behavior of the ensemble. This generalizes known results for LDPC codes and GLDPC codes, and also generalizes the corresponding connection with the stability condition over the BEC.

\appendices

\section{Proof of Lemma~\ref{lemma:O2_term_check}}
\label{app:proof_lemma}
Consider any $\bldbeta^{(t)}$ which satisfies the optimization constraints~(\ref{eq:sum_epsilont_constraint})--(\ref{eq:nonnegative_betat_constraint}). From constraint~(\ref{eq:sum_epsilont_constraint}), $\delta$ small implies that $\epsilon_t$ is small for every $t \in I_c$. From constraint (\ref{eq:sum_t_betat_constraint}) we conclude that $\beta_{i}^{(t)}$ is small for every $t \in I_c$, $i \in U_t^{-}$, and so $\beta_{0}^{(t)}$ is close to $1$ for all $t \in I_c$. Formally, for any $t \in I_c$ the term in the sum over $i \in U_t$ in (\ref{eq:W}) corresponding to $i = 0$ may be written as (here we use (\ref{eq:betat_sum_constraint}), and the Taylor series of
$\log\left(1-x\right)$ around $x=0$)
\begin{eqnarray*}
\beta^{(t)}_{0} \log \left( \frac{1}{\beta^{(t)}_{0}} \right) & = & \Big( \sum_{i \in U_t^{-}} \beta^{(t)}_{i} - 1 \Big) \log \Big( 1 - \sum_{i \in U_t^{-}} \beta^{(t)}_{i} \Big) \\
& = & \Big( \sum_{i \in U_t^{-}} \beta^{(t)}_{i} - 1 \Big) \Bigg(- \sum_{i \in U_t^{-}} \beta^{(t)}_{i} + O \Big( \Big( \sum_{i \in U_t^{-}} \beta^{(t)}_{i} \Big)^2 \Big) \Bigg) \\
& = & \sum_{i \in U_t^{-}} \beta^{(t)}_{i} + O \Big( \Big( \sum_{i \in U_t^{-}} \beta^{(t)}_{i} \Big)^2 \Big) 
\end{eqnarray*}
Therefore we have
\[
F_t(\bldbeta^{(t)}) = O \Big( \Big( \sum_{i \in U_t^{-}} \beta^{(t)}_{i} \Big)^2 \Big)
\]
i.e.
\begin{equation} 
\left| F_t(\bldbeta^{(t)}) \right| \le k_t \Big( \sum_{i \in U_t^{-}} \beta^{(t)}_{i} \Big)^2
\label{eq:F_t_bound}
\end{equation}
for some $k_t > 0$ independent of $\{ \beta^{(t)}_{i} \}_{i \in U_t^{-}}$. It follows that
\begin{equation}
\left| \sum_{t \in I_c} \gamma_t F_t(\bldbeta^{(t)}) \right| \le \sum_{t \in I_c} \gamma_t \left| F_t(\bldbeta^{(t)}) \right| \le \sum_{t \in I_c} \gamma'_t \Big( \sum_{i \in U_t^{-}} \beta^{(t)}_{i} \Big)^2
\label{eq:F_t_ineq_1}
\end{equation} 
where $\gamma'_t = k_t \gamma_t$ for each $t \in I_c$. Also, by (\ref{eq:sum_t_betat_constraint}) we have $\sum_{i \in U_t^{-}} \beta^{(t)}_{i} \le \epsilon_t / \gamma_t$ and therefore
\begin{equation}
\sum_{t \in I_c} \gamma'_t \Big( \sum_{i \in U_t^{-}} \beta^{(t)}_{i} \Big)^2 \le \sum_{t \in I_c} \left( \frac{\gamma'_t}{\gamma_t^2} \right) \epsilon_t^2
\label{eq:F_t_ineq_2}
\end{equation} 
Denote $\gamma = \max_{t \in I_c} \{ \gamma'_t / \gamma_t^2 \}$; then, combining (\ref{eq:F_t_ineq_1}) and (\ref{eq:F_t_ineq_2}), 
\[
\left| \sum_{t \in I_c} \gamma_t F_t(\bldbeta^{(t)}) \right| \le \gamma \sum_{t \in I_c} \epsilon_t^2 < \gamma \left( \sum_{t \in I_c} \epsilon_t \right) ^2 = \gamma \delta^2
\]
and thus the expression $\sum_{t \in I_c} \gamma_t F_t(\bldbeta^{(t)})$ is $O(\delta^2)$, as desired.

\section{Growth Rate of the Stopping Set Size Distribution}
\label{app:stopping_sets}
This appendix illustrates how the approach developed to analyze the growth rate of the weight distribution of D-GLDPC codes can also be used to analyze the growth rate of the stopping set size distribution. The concept of stopping set was introduced in \cite{di02:finite} within the context of iterative decoding of LDPC codes over the \ac{BEC}. A stopping set of an LDPC code is defined as any subset of the \acp{VN} such that if a \ac{CN} is connected to it, it is connected to it at least twice. Over the \ac{BEC}, stopping sets under iterative decoding play the same role as codewords under maximum likelihood decoding.

Stopping sets can also be defined within the context of D-GLDPC codes over the \ac{BEC}. In contrast to the case of LDPC codes however, here the definition of stopping set is not unique as it depends on the decoding algorithm used at the \acp{VN} and \acp{CN} to locally recover from erasures. In the following, we assume \ac{MAP} erasure decoding is used at both the \acp{VN} and the \acp{CN}.

Consider an $(n,k)$ linear block code and a generator matrix $\mathbf{G}$ for this code. Moreover, consider a $k$-bit information word $\mathbf{u}$ containing erasures. Encoding of this information word produces an $n$-bit word $\mathbf{x}$ containing erasures, where the non-erasure encoded bits are those depending only on the non-erasure information bits (through the relationship $\mathbf{x}=\mathbf{u}\mathbf{G}$). We say that the erasure pattern on the encoded bits is \emph{induced} by the erasure pattern on the information bits.

Consider an $(s,h)$ \ac{CN} of a D-GLDPC code over the \ac{BEC}, and let $\mathbf{G}$ denote a generator matrix for this \ac{CN}. A \emph{local stopping set} for this \ac{CN} is a subset of the local code bits such that, if all of these bits are erased, \ac{MAP} decoding cannot recover any of these bits. This occurs if and only if each column of $\mathbf{G}$ corresponding to erased bits is linearly independent of the columns of $\mathbf{G}$ corresponding to the non-erased bits\footnote{Note that the definition of local stopping set is independent of the particular generator matrix chosen for the \ac{CN}.}. The \emph{size} of the local stopping set for the \ac{CN} is equal to the number of erased local code bits. 

Next consider a $(q,k)$ \ac{VN} of a D-GLDPC code over the \ac{BEC}, and let $\mathbf{G}$ denote the generator matrix for this \ac{VN}, i.e. $\mathbf{G}$ expresses the relationship between the local information word and the local codeword whose bits are associated with the Tanner graph edges. A \emph{local stopping set} for this \ac{VN} is a subset of the local information bits, together with a subset of the local code bits, such that if all of these bits are erased, \ac{MAP} decoding at the \ac{VN} cannot recover any of these bits. Since \ac{MAP} decoding over the \ac{BEC} consists of running Gaussian elimination on the $(k \times (q+k))$ matrix $\mathbf{G}'=[\mathbf{G} \; | \; \mathbf{I}_k]$, where $\mathbf{I}_k$ is the $(k \times k)$ identity matrix, this occurs if and only if each column of $\mathbf{G}'$ corresponding to an erased bit is linearly independent of the columns of $\mathbf{G}'$ corresponding to the non-erased bits. The \emph{split size} of a \ac{VN} stopping set is equal to $(u,v)$ where there are $u$ erasures among the local information bits, and $v$ erasures among the local code bits.

The concepts of \emph{assignment} and \emph{split assignment} remain valid also within the present context of stopping sets. However we redefine the concepts of check-valid assignment, check-valid split assignment and variable-valid split assignment as follows.
\medskip
\begin{definition}\label{def:assignment_stopping_set}
An assignment is said to be \emph{check-valid} if the following condition holds: supposing that each edge of the assignment carries an erasure and each of the other edges carries a non-erasure, each \ac{CN} recognizes a local stopping set. 
\end{definition}
\medskip
\begin{definition}\label{def:split_assignment_stopping_set}
A split assignment is said to be \emph{check-valid} if its assignment is check-valid. A split assignment is said to be \emph{variable-valid} if the following holds. Supposing that each edge of its assignment carries an erasure and each of the other edges carries a non-erasure, and supposing that each D-GLDPC code bit in the codeword assignment is set to an erasure and each of the other code bits to a non-erasure, each \ac{VN} recognizes a local stopping set, where the erasure pattern on its local code bits is that induced by the erasure pattern on its local information bits.
\end{definition}
\medskip
A stopping set of a D-GLDPC code may be defined as a codeword assignment such that the split assignment formed by the codeword assignment and the corresponding induced assignment is both variable-valid and check-valid. The \emph{size} of such a stopping set is equal to the number of elements in the codeword assignment. The growth rate of the stopping set size distribution is defined as in \eqref{eq:growth_rate_result}, where in this context $N_{\alpha n}$ denotes the number of stopping sets of size $\alpha n$. 

We define the stopping set size enumerating polynomial for \ac{CN} type $t \in I_c$ by 
$$\varPhi^{(t)}(x)=1+\sum_{u=r_t}^{s_t}\varphi^{(t)}_u x^u$$ 
where $\varphi^{(t)}_u$ denotes the number of local stopping sets of size $u$ for \acp{CN} of type $t$. Letting $\mathbf{G}_t$ denote any generator matrix for \ac{CN} type $t$, note that $\varphi^{(t)}_u$ is equal to the number of ways of choosing $u$ columns of $\mathbf{G}_t$ such that each of the selected columns is linearly independent of the $s_t-u$ non-selected columns. 

We also define the bivariate stopping set split size enumerating polynomial for \ac{VN} type $t \in I_v$ by 
$$\varTheta^{(t)}(x,y)=1+\sum_{u=1}^{k_t}\sum_{v=p_t}^{q_t}\vartheta^{(t)}_{u,v} x^u y^v$$ 
where $\vartheta^{(t)}_{u,v}$ denotes the number of local stopping sets of split size $(u,v)$ for \acp{VN} of type $t$. Letting $\mathbf{G}_t$ denote the generator matrix for \ac{VN} type $t$, note that $\vartheta^{(t)}_{u,v}$ is the number of ways of choosing $u$ columns of $\mathbf{I}_{k_t}$ and $v$ columns of $\mathbf{G}_t$ such that each of the selected columns is linearly independent of the $k_t + q_t - v - u$ non-selected columns of $\mathbf{G}_t'=[\mathbf{G}_t \; | \; \mathbf{I}_{k_t}]$. 

With these definitions in place, the analog of Theorem \ref{thm:growth_rate} may be developed in an identical manner for the growth rate of the stopping set size distribution. The proof is identical to that developed in Section~\ref{section:proof_of_main_result} for the growth rate of the weight distribution, with the stopping set size enumerators in place of the weight enumerators, the new definitions of check- and variable-validity, and ``stopping set size'' in place of ``codeword weight''. In particular, we have the following result for the case $r=p=2$, which is the most important case in practice.
\medskip
\begin{theorem}\label{theorem:growth_rate_stopping_sets}
Consider a D-GLPDC code ensemble $\cM_n$ satisfying $r=p=2$. Assume the code is transmitted over the \ac{BEC}, and it is decoded via iterative decoding with \ac{MAP} erasure decoding at the \acp{VN} and \acp{CN}. The growth rate of the stopping set size distribution is given by
\begin{equation}
G(\alpha) = \alpha \log \left[ \frac{1}{P^{-1}(1/C)} \right] + O(\alpha^{3/2})
\label{eq:growth_rate_stopping_sets}
\end{equation}
where the polynomial $P(x)$ and the parameter $C$ are given by (\ref{eq:Pofx_definition}) and (\ref{eq:C_definition_revisited}) respectively.
\end{theorem}
\medskip
The proof of this result is completed by observing that, if $r=p=2$, we have $\varphi^{(t)}_{2} = A^{(t)}_2$ for each $t \in X_c$ and $\vartheta^{(t)}_{u,2}=B^{(t)}_{u,2}$ for each $t \in X_v$, $u=1,2,\cdots, k_t$. Note that, in contrast to (\ref{eq:growth_rate_case_1}), the error term is always $O(\alpha^{3/2})$ in (\ref{eq:growth_rate_stopping_sets}) because in the context of stopping sets, $p=r=2$ implies $\bar{p}=\bar{r}=3$.

Finally, note that \eqref{eq:growth_rate_stopping_sets} generalizes to D-GLDPC codes the expression $G(\alpha)=\alpha \log (\lambda'(0)\rho'(1)) + o(\alpha)$ obtained in \cite{orlitsky05:stopping} for the growth rate of the stopping set size distribution of LDPC codes.

\section*{Acknowledgments}
The authors would like to thank A. Barg, M. Lentmaier and V. Skachek for helpful discussions. They would also like to thank the anonymous reviewers for their comments which helped to significantly improve an earlier draft of the paper. 


\bibliographystyle{IEEEtran}

\end{document}